\documentclass[aps,prx,twocolumn,superscriptaddress,groupedaddress,nofootinbib,showpacs]{revtex4}  
\usepackage{graphicx}  
\usepackage{dcolumn}   
\usepackage{bm}        
\usepackage{amssymb}   
\usepackage{mathtools}
\usepackage[caption=false]{subfig}
\usepackage{floatrow}
\usepackage{float}
\usepackage[utf8x]{inputenc}
\usepackage[makeroom]{cancel}
\usepackage{capt-of}
\usepackage{chemscheme}
\usepackage[detect-all]{siunitx}
\usepackage{titletoc}
\newfloatcommand{capbtabbox}{table}[][\FBwidth]

\begin{document}

\makeatletter
\def\scheme{\@float{scheme}}
\let\endscheme\endfigure
\makeatother



\title{Modeling Evolution of Crosstalk in Noisy Signal Transduction Networks}
\author{Ammar Tareen}
\affiliation{Clark University, Department of Physics, Worcester, MA 01610}
\author{Ned S. Wingreen}
\email{wingreen@princeton.edu}
\affiliation{Lewis-Sigler Institute for Integrative Genomics, Carl Icahn Laboratory, Washington Road, Princeton, NJ 08544}
\author{Ranjan Mukhopadhyay}
\email{ranjan@clarku.edu}
\affiliation{Clark University, Department of Physics, Worcester, MA 01610}

\date{\today}

\begin{abstract}
Signal transduction networks can form highly interconnected systems within cells due to network crosstalk, the sharing of input signals between multiple downstream responses. To better understand the evolutionary design principles underlying such networks, we study the evolution of crosstalk and the emergence of specificity for two parallel signaling pathways that arise via gene duplication and are subsequently allowed to diverge. We focus on a sequence based evolutionary algorithm  and evolve the network based on two physically motivated fitness functions related to information transmission. Surprisingly, we find that the two fitness functions lead to very different evolutionary outcomes, one with a high degree of crosstalk and the other without.
\end{abstract}

\maketitle


\section{Introduction}

Signaling networks have evolved to transduce external and internal information to inform critical cellular decisions such as growth, differentiation, directional motion, metabolic transitions, and apoptosis \cite{Jordan}. These networks can form highly interconnected systems within cells due to network crosstalk, the sharing of input signals among multiple canonical pathways. Crosstalk between pathways accounts for many of the complex behaviors exhibited by signaling networks \cite{crosstalk_experimental_exist_1,crosstalk_experimental_exist_2,crosstalk_experimental_exist_3,CT_experiment_4,Deeds-Eukaryote,ashok}. How did such complex interconnected networks evolve and what constraints did the dynamics of evolution place on their architecture? Can we understand the resulting degree of crosstalk in terms of optimization of fitness associated with the accurate transmission of information? 

In cells, there exist examples of both high degrees of crosstalk and high degrees of specificity. As an example of crosstalk, studies have shown interactions between the IGF-I and the TGF-$\beta$ pathways, where in the Hep3B human hepatoma cell line, IGF-I and insulin were each shown to block TGF-$\beta$ induced apoptosis, via a PI3-kinase/Akt dependent pathway \cite{CT_example_1_IGFITGFBeta_Interaction}. In another example of crosstalk, cyclic AMP helps regulate cell proliferation by interacting with the mitogen-activated protein (MAP) kinase pathway \cite{CT_example_2_camp_mapk_interaction}. More examples can be found in \cite{CT_example_3_integrin_signaling,CT_example_4_hunter,CT_example_5_yan,CT_example_6_nishi}. On the other hand, two-component signaling systems, which form the dominant signaling modality in bacteria, exhibit a high degree of pathway isolation and therefore a high degree of specificity \cite{Deeds-Prokaryote}. Examples of specificity in signaling are found in \cite{specificity_example_1_Sharad,specificity_example_2_Behar,specificity_example_3_Bardwell,specificity_example_4_Bardwell,specificity_example_5_Flatauer,specificity_example_6_Dard}. Indeed, undesirable crosstalk underlies many pathological conditions in higher organisms \cite{Muller,Shi,Kalaitzidis}.

Understanding the evolutionary drive toward pathway specificity or crosstalk is a fundamental problem in signal transduction. However, modeling the evolution of crosstalk in signaling networks is challenging because evolutionary processes are governed by changes at the genotypic level, whereas selection occurs at the phenotypic level \cite{evolution} and the mapping  between genotype and phenotype is generally poorly understood. Currently, much of the theory related to evolution of signal transduction networks focuses on changes at the phenotypic level (e.g. changing protein interactions directly) \cite{Mobashir,soyer}. In this paper we adapt a sequence-based evolutionary model due to Zulfikar \textit{et al}. \cite{zulfikar} that allows us to map from sequence space (genotype) to rate constant space (phenotype). For the first time, we apply this model to signal transduction in order to better understand the evolution of crosstalk and the emergence of specificity. 

\indent New signaling pathways can enter the genome via gene duplication and subsequent divergence \cite{geneDup_Prince}. Therefore, in this paper, we consider two parallel pathways that arise via gene duplication but then are allowed to diverge. We evolve our network using two biologically motivated fitness functions related to the transmission of information. For the first fitness function, we focus on a system which may have evolved to transmit the total information content along the signaling network. Drawing from Shannon's work on communication theory \cite{shannon}, a suitable choice of fitness for this scenario is the total mutual information, denoted by $\mathrm{MI_{total}}$. For the second fitness function, we consider a system where inputs transmitted through their cognate signaling pathways lead to distinct responses. A natural choice of fitness function for this scenario is the sum of the mutual informations of individual pathways, denoted by $\mathrm{MI_{sum}}$. 
We find that the two fitness functions lead to very different evolutionary outcomes. In particular, evolution retains a high degree of crosstalk for the case of $\mathrm{MI_{total}}$ while leading to high specificity for $\mathrm{MI_{sum}}$. 

\section{Evolutionary Model}

\captionsetup[subfigure]{singlelinecheck=false,justification=raggedright,position=top}
\begin{figure}[htp]
\begin{center}
\subfloat[]{%
  \includegraphics[clip,width=0.75\columnwidth]{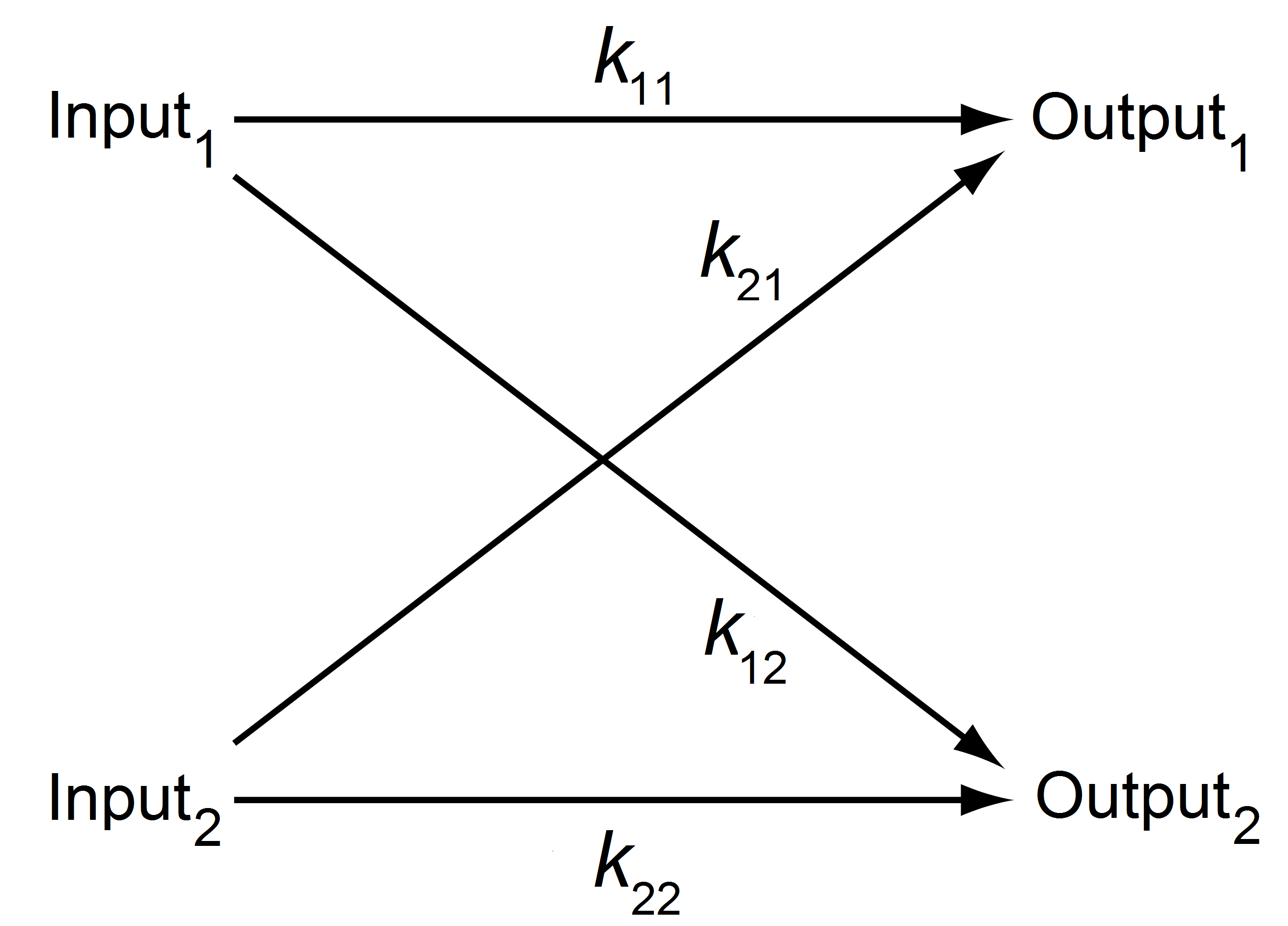}%
  \label{2D_System_EV}
}

\subfloat[]{%
  \includegraphics[clip,width=0.75\columnwidth]{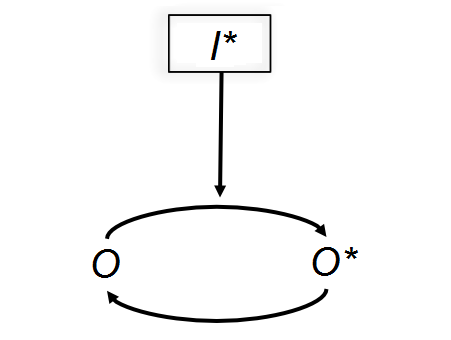}%
  \label{schematics1}
}
\end{center}
\caption{Signaling schematics. (a) Signal transduction network showing both direct and crosstalk pathways along with their associated reaction rate constants. (b) Schematic illustration of a signaling pathway. $I^*$ represents input and $O^*$ represents output.}
\label{schematics}
\end{figure}

In our model of a signaling pathway, we assume two layers of proteins that represent an input-output process. The first layer corresponds to a set of proteins (e.g. cell surface receptors or protein kinases) that become activated by an extracellular signal (e.g. a ligand); the activated fraction of these proteins represents the input. These input proteins, in their active form, can  activate a second layer of proteins whose activated fraction represents the output. To study information transmission in this system (see Fig. \ref{schematics}), we employ the chemical Langevin equation, which approximately models the stochastic dynamical behavior of a well-stirred mixture of molecular species that chemically interact \cite{Gillespie}:


\begin{equation}
\begin{split}
\frac{d{O^*_j}}{dt} = \overbrace{\sum _{i,j}k_{ij} I_i O_j}^\text{${O_{j,\textrm{activation}}}$}   \hspace{0.3 cm}- & \overbrace{\alpha {O^*_j}}^\text{${O^*_{j,\textrm{deactivation}}}$}+ \\
& \sqrt{\frac{      \sum_{i,j}k_{ij} I_i O_j+\alpha {O^*_j}}{V}}\xi_j(t). 
\end{split}
\label{chemical_langevin}
\end{equation}
$I_i$ is the strength of input $i$, $O^*_j$
is the concentration of activated output protein $j$ (aka the output), and $O_j$ is the inactive concentration, with the total concentration of output protein held fixed i.e. $O_\mathrm{tot}= O_j+O^*_j$. We assume  a background deactivation rate of $\alpha = 1$ and $O_\mathrm{tot}=1$, which define our units of time and volume. $V$ represents the volume of the system, and controls the level of noise. The factors $k_{ij}$ are reaction rate constants. $\xi_j$ are temporally uncorrelated, statistically independent Gaussian white-noise terms whose coefficients are defined by the square root of the sum of the activation and deactivation rates, see \cite{gonze}. The $\xi_j$ have zero mean  $< \xi_j(t)> = 0$, and are delta-correlated in time $< \xi_j(t) \xi_j({t}^{\prime})> = \delta(t-t^{\prime})$.

Crosstalk in signaling networks results from the intrinsic promiscuity of protein-protein interactions \cite{ladbury}. Protein-protein interaction strengths are generally determined by amino-acid-residue interactions at specific molecular interfaces. Moreover, it has been estimated that greater than 90\% of protein interaction interfaces are planar with the dominant contribution coming from hydrophobic interactions. For simplicity, we therefore assume that input proteins possess an out-face and output proteins possess an in-face which form a pair of interaction interfaces; we associate a binary sequence, $\vec{\sigma}_{in,out}$, of hydrophobic residues (1s) and hydrophilic residues (0s) to each interface. The interaction strength between a protein (denoted by index $i$) and its target (denoted by index $j$) is determined by the interaction energy $E_{ij} = \epsilon \vec{\sigma}_{out}^{i} \cdot \vec{\sigma}_{in}^{j}$ between the out-face of the input protein and the in-face of the output protein. $\epsilon$ represents the effective interaction energy between two hydrophobic residues. (All energies are expressed in units of the thermal energy $k_B T$.) The reaction rate is then given by

\begin{equation}
k_{ij} = {\frac{k_0}{1 + \exp[-(E_{ij} - E_0)]}},
\label{kil}
\end{equation}
where $E_0$ plays the role of a threshold energy, e.g. accounting for the loss of entropy due to binding. In our calculations we varied $k_0$ between \numrange{1}{20}, $\epsilon$ between \numrange{0.2}{0.6}, and $V$ between \numrange{1}{10}. We set $E_0 = 5$, and we took the length of each sequence representing an interface to be $M = 25$. These interaction parameters were chosen to provide a large range (\numrange{0.006}{20}) for the possible rate constants $k_{ij}$ as a function of sequence and to keep the background deactivation rates small compared to the highest activation rates.

For our evolutionary scheme, we assume a population sufficiently small that each new mutation is either fixed or entirely lost \cite{moran, nowak}. We consider only point mutations - namely replacing a randomly chosen hydrophobic residue (1) in the in- or out-face of one protein by a hydrophilic residue (0), or vice versa. In this study, mutations are accepted if and only if they produce a fitness that is greater than or equal to the current fitness. In this work, we studied two fitness functions based on the mutual information between the inputs and outputs of our system, with MI defined as \cite{shannon}:

\begin{equation}
\mathrm{MI}(I;O^*) = \iint P(I,O^*) \log \frac{P(I,O^*)}{P(I)P(O^*)} dI d O^* ,
\label{MI}
\end{equation}
where $P$ always represents a probability distribution function. For simplicity, we chose the input probability distribution $P(I)$ to be Gaussian. The mutual information (MI) of two random variables is a measure of the mutual dependence between the two variables. The two fitness functions that we studied here are based on the above general definition of mutual information and can be expressed as follows:

\begin{gather}
   \mathrm{MI}_{\mathrm{total}} =   \mathrm{MI}(I_1,I_2; O^*_1,O^*_2),	\nonumber
\\
  \mathrm{MI}_{\mathrm{sum}} = \mathrm{MI}(I_1;O^*_1)+\mathrm{MI}(I_2;O^*_2).
  \label{Fitness_Functions}
  \end{gather}
Qualitatively, $\mathrm{MI_{total}}$ represents the fitness for a system which evolves to transmit the total information content via the entire signaling network, whereas $\mathrm{MI_{sum}}$ represents fitness for a system where inputs transmitted through their cognate signaling pathways lead to distinct responses.

\section{Phenotypic Fitness Landscapes}

To implement the above evolutionary model, we must be able to calculate mutual information. We use the Fokker-Planck (FP) equation \cite{Risken} corresponding to our Langevin equation (Eq. \ref{chemical_langevin}) to calculate the probability distributions appearing in the MI (Eq. \ref{MI}). We first consider the simpler case of a one-input, one-output system to develop tools to address multiple input-output systems with crosstalk.

\subsection{Single Input and Output: No Crosstalk}

\begin{figure}[htp]
\includegraphics[scale=0.3]{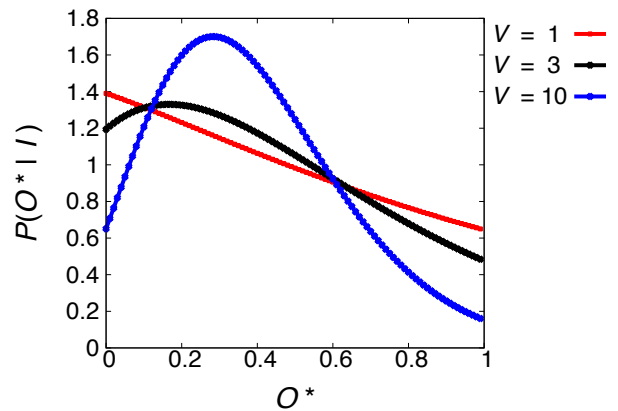}%
\caption{Conditional output probability distribution. The conditional output probability distribution shown for $I =\frac{1}{2}$ and $k_{11} = 1$ for several values of system volume $V$.}
 \label{condoutput1}
\end{figure}

For a one-input, one-output system, as shown schematically in Fig. \ref{schematics1}, the Langevin equation can be written as a deterministic part $A$ and a stochastic part $B$:

\begin{equation}
\frac{dO^*}{dt} = A(O^*,t)+B(O^*,t)\xi(t),
\end{equation}
where $A$ and $B$ are defined as follows.

\begin{gather}
   A(O^*,t) = k_{11}IO-\alpha O^*,	\nonumber 
\\
  B(O^*,t) = \sqrt{\frac{k_{11}IO +\alpha O^*}{V}}.
  \label{strengthofnoise}
  \end{gather}  
The resulting FP equation (in the It$\mathrm{\hat{o}}$ formulation \cite{vanKampen}) is

\begin{equation}
\frac{\partial P}{\partial t} = -\frac{\partial }{\partial O^*} \Big \{ A(O^*,t) P \Big\}+
 \frac{1}{2}\frac{\partial^2 }{\partial {O^*}^2} \Big\{ B^2(O^*,t) P \Big\}.
\label{FP1}
\end{equation}
Note that Eq. \ref{FP1} has the form of a continuity equation for probability
\begin{equation}
\frac{\partial P(O^*,t)}{\partial t} + \frac{\partial J(O^*,t)}{\partial O^*} = 0,
\label{J1}
\end{equation}
where $J = \frac{\partial }{\partial O^*}  (AP - \frac{1}{2} (B^2 P))$ can be viewed as a probability current. The steady-state solution of the FP equation corresponds to a constant value of $J$. Imposing the boundary conditions $J=0$ at $O^* = 0$ and at $O^*=1$ then implies that $J=0$ everywhere. The solution of the steady-state FP equation for zero-probability-current boundary conditions can be written as \cite{supp} 


\begin{equation}
\resizebox{1.0\hsize}{!}{$
P(O^* | I,k_{11})  = N {e^{\frac{-2 V O^* (I k_{11}+\alpha)}{\alpha-Ik_{11}}}} {\Big[1+\frac{(\alpha-I k_{11})O^*)}{I k_{11} O_{\textrm{tot}}}\Big]}^{{\frac{4 I k_{11} O_{\textrm{tot}} \alpha V}{(\alpha-I k_{11})^2}}-1},
$}
\label{P_y_given_x}
\end{equation}
where $N$ is a normalization constant. Note that while this conditional output probability distribution is peaked for $V=2$ or higher, it does not resemble a Gaussian distribution even at reasonably large values of $V$ (Fig. \ref{condoutput1}). Additionally, it might appear that the RHS of Eq. \ref{P_y_given_x} approaches $\infty$ as $\alpha \rightarrow I k_{11}$; however setting $\delta = \alpha - I k_{11}$ and Taylor expanding around $\delta = 0$, we find that the divergent terms cancel \cite{supp}.

\begin{figure}[H]
\begin{center}
\includegraphics[scale=0.265]{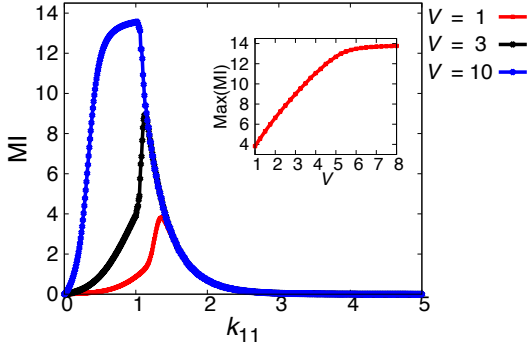}
\caption{Mutual Information versus $k_{11}$ shown for several values of system volume $V$. The inset shows the maximum value of mutual information as a function of system volume V. The input probability distribution, $P(I)$, is chosen to be a Gaussian (mean $\mu =  0.5 $ and standard deviation $\sigma = 0.1$).}
\label{MI_Several_V_fixed_C}
\end{center}
\end{figure}

We can determine the output probability $P(O^*)$ by numerically integrating the conditional output probability over the input distribution as follows,
\begin{equation}
P(O^*) =  \int  P(O^*|I)P(I)dI.
\label{p-out}
\end{equation}
We obtain the mutual information as a function of $k_{11}$, as shown in Fig.  \ref{MI_Several_V_fixed_C}. The mutual information is nearly zero both at very small values of $k_{11}$ because of low activation and at very large values of $k_{11}$ because of saturated output. The inset in Fig. \ref{MI_Several_V_fixed_C} shows the maximum value of mutual information as a function of system volume $V$; the maximum mutual information starts flattening out for $V > 5$.

\subsection{Duplicated Inputs and Outputs: Nonzero Crosstalk}

We now extend the one-input, one-output system to two inputs and two outputs, and allow for crosstalk. The Langevin equations for this system are

\begin{equation}
\begin{split}
\frac{d{O^*_1}}{dt} = & k_{11} I_1 O_1 + k_{21} I_2 O_1 - \alpha {O^*_1}  \\
& + \sqrt{\frac{k_{11} I_1 O_1  + k_{21} I_2 O_1 + \alpha {O_1}^*}{V}}\xi_1(t),
\end{split}
\label{SP_1}
\end{equation}

\begin{equation}
\begin{split}
\frac{d{O^*_2}}{dt} = & k_{12} I_1 O_2  + k_{22} I_2 O_2 - \alpha {O^*_2} \\
&+ \sqrt{\frac{k_{12} I_1 O_2  + k_{22}I_2 O_2 +\alpha {O_2}^*}{V}}\xi_2(t).
\end{split}
\label{SP_2}
\end{equation}
Compactly, the multivariate Langevin equation can be written as

\begin{equation}
\frac{d O^*_i}{dt} = A_i(\textbf{O}^*,t)+\sum\limits_{{k=1} }^2 B_{ik}(\textbf{O}^*,t)\xi_k(t),
\end{equation}
where the index $i$ can take on the values \{1,2\}, and where the functions $A$ and $B$ are represented by the first and second terms, respectively, in Eq. \ref{SP_1} and Eq. \ref{SP_2}. The resulting FP equation for the joint probability distribution $P(O_1^*, O_2^*,t)$ is \cite{Garcia-Palacios,multivariateFP}:

\begin{equation}
\begin{split}
\frac{\partial P}{\partial t} = -\sum_i  \frac{\partial }{\partial O^*_i} \Big \{ A_i(\textbf{O}^*,t)  P \Big\} +  \\
\frac{1}{2} \sum_{ij} \frac{\partial^2 }{\partial O^*_i \partial O^*_j} \Big\{ \Big[ \sum_k B_{ik}(\textbf{O}^*,t)B_{jk}(\textbf{O}^*,t) \Big] P \Big\}.
\end{split}
\label{FPE_ii}
\end{equation}
The steady-state solution that satisfies the zero-probability-current boundary conditions for Eq. \ref{FPE_ii} is \cite{supp} 


\begin{equation}
\resizebox{1.0\hsize}{!}{$
P(O^*_i|I_1,I_2) = N e^{[-2 V O^*_i \frac{\alpha R_i^* +1}{\alpha R_i^* -1}]} [1+\frac{(\alpha R_i^* -1)O^*_i}{R_i R_i^*}]^{\frac{4 V R_i {R_i^*}^2 \alpha}{(\alpha R_i^* -1)^2} -1}.
$}
\label{P_2D_Cond_output}
\end{equation}
For notational convenience we have introduced modified rates defined as follows:

\begin{equation}
R_1 \equiv O_{\textrm{tot},1} (k_{11} I_1+k_{21} I_2),	\nonumber
\qquad
R_1^* \equiv O^*_1 (k_{11} I_1+k_{21} I_2).
\qquad
\label{R1T1}
\end{equation}

\begin{equation}
R_2 \equiv O_{\textrm{tot},2} (k_{12} I_1+k_{22} I_2),
\qquad
R_2^* \equiv O^*_2 (k_{12} I_1+k_{22} I_2).
\qquad
\label{R2T2}
\end{equation}

\begin{figure}
\includegraphics[scale = 0.3]{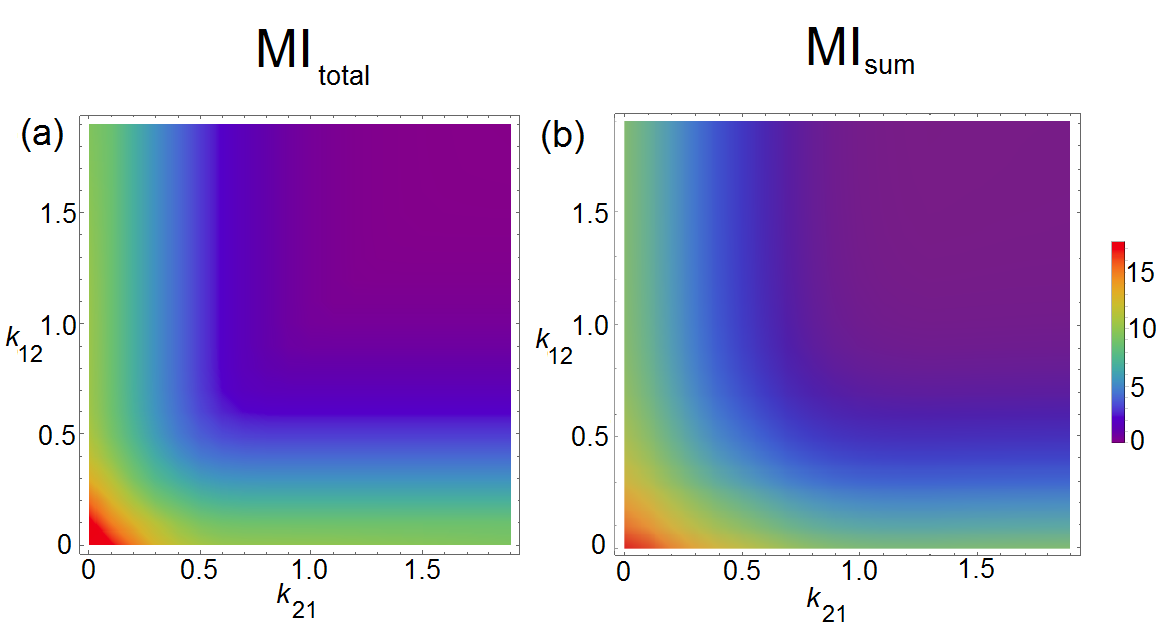}
\caption{Phenotypic fitness landscapes. (a). $\mathrm{MI_{total}}$ and (b) $\mathrm{MI_{sum}}$ versus crosstalk rate constants $k_{21}$ and $k_{12}$, with $k_{11}=k_{22}=1$, for $V=3$.  
}
\label{MI_2D_Density}
\end{figure}

Having obtained the conditional probabilities, we can numerically obtain the two fitness functions using Eqs. \ref{MI} and \ref{Fitness_Functions}. For $V=3$, if we set $k_{11} = k_{22} = 1$ (i.e. corresponding to values of these rate constants close to the optimum of MI for a single pathway, as seen in Fig. \ref{MI_Several_V_fixed_C}), then we can depict the density plots of fitness versus. crosstalk, as in Fig. \ref{MI_2D_Density}, and observe that the optima for both fitness functions occur at zero crosstalk (larger volumes yield qualitatively similar landscapes, see \cite{supp} for a calculation with $V=10$). Both fitness landscapes look similar and both have a fitness maximum at zero crosstalk. However, Fig. \ref{MI_2D_Density} provides only a slice through parameter space. How might an evolving system explore the full space? To answer this question we take an evolutionary approach.

\section{Evolution of Crosstalk}



\begin{figure}
\begin{center}
\includegraphics[scale = 0.275]{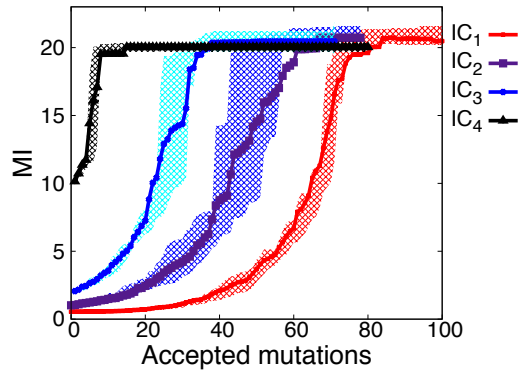}
\caption{Evolution of fitness versus accepted mutations. Four different initial conditions are shown (labelled \lq{IC}\rq \hspace{0.2 mm} on the legend) \cite{supp}; solid curves represent fitness averaged over 100 simulations while shaded curves represent 25-75 percentiles from each of the simulations at every accepted mutation. Mutations are accepted only if the new fitness is greater than or equal to the previous one. A higher number of 1s in the initial strings leads to a higher starting value in MI. $k_0 = 20$, $E_0 = 5$, $V=3$, $\epsilon=0.2$.}
\label{MI_greedy_optimization}
\end{center}
\end{figure}

The fitness landscapes in Fig. \ref{MI_2D_Density} are obtained by setting direct rate constants $k_{11} = k_{22} = 1$ (for $V=3$) and by sweeping over values of the crosstalk rate constants $k_{12}, k_{21}$. This choice of direct rate constants is motivated by the value of the rate constant in Fig. \ref{MI_Several_V_fixed_C}, for which $k_{11}=1$ produces a maximum. However, this one slice through the four-dimensional space of rate constants cannot capture the full fitness landscape. Moreover, we expect biologically that random mutations will change more than one rate constant. Motivated by these considerations, we therefore implemented an evolutionary algorithm in which we make random mutations to the binary strings that determine the rate constants (Eq. \ref{kil}), and accept mutations if and only if they produce a fitness that is greater than or equal to the current fitness (Eq. \ref{Fitness_Functions}). The initial state of the system corresponds to duplicated pathways where all the rate constants $k_{ij}$ are equal (e.g. for all strings initialized to zero and $\epsilon=0.2$, $k_{ij} \approx 0.1$). Fig. \ref{MI_greedy_optimization} shows some sample runs of the evolutionary algorithm for a few different choices of initial conditions; each solid curve represents the average fitness for one hundred runs for a specific set of initial strings, while the shaded regions indicate the 25-75 fitness percentiles at that particular number of accepted mutations over all trajectories. Fig. \ref{MI_greedy_optimization} shows results for $\mathrm{MI_{total}}$, however the results for $\mathrm{MI_{sum}}$ are the same qualitatively. We can see that the final values of the rate constants do not depend critically on our choice of initial strings. 

\begin{figure}[htp!]
\begin{center}
\subfloat[]{%
  \includegraphics[scale=0.255]{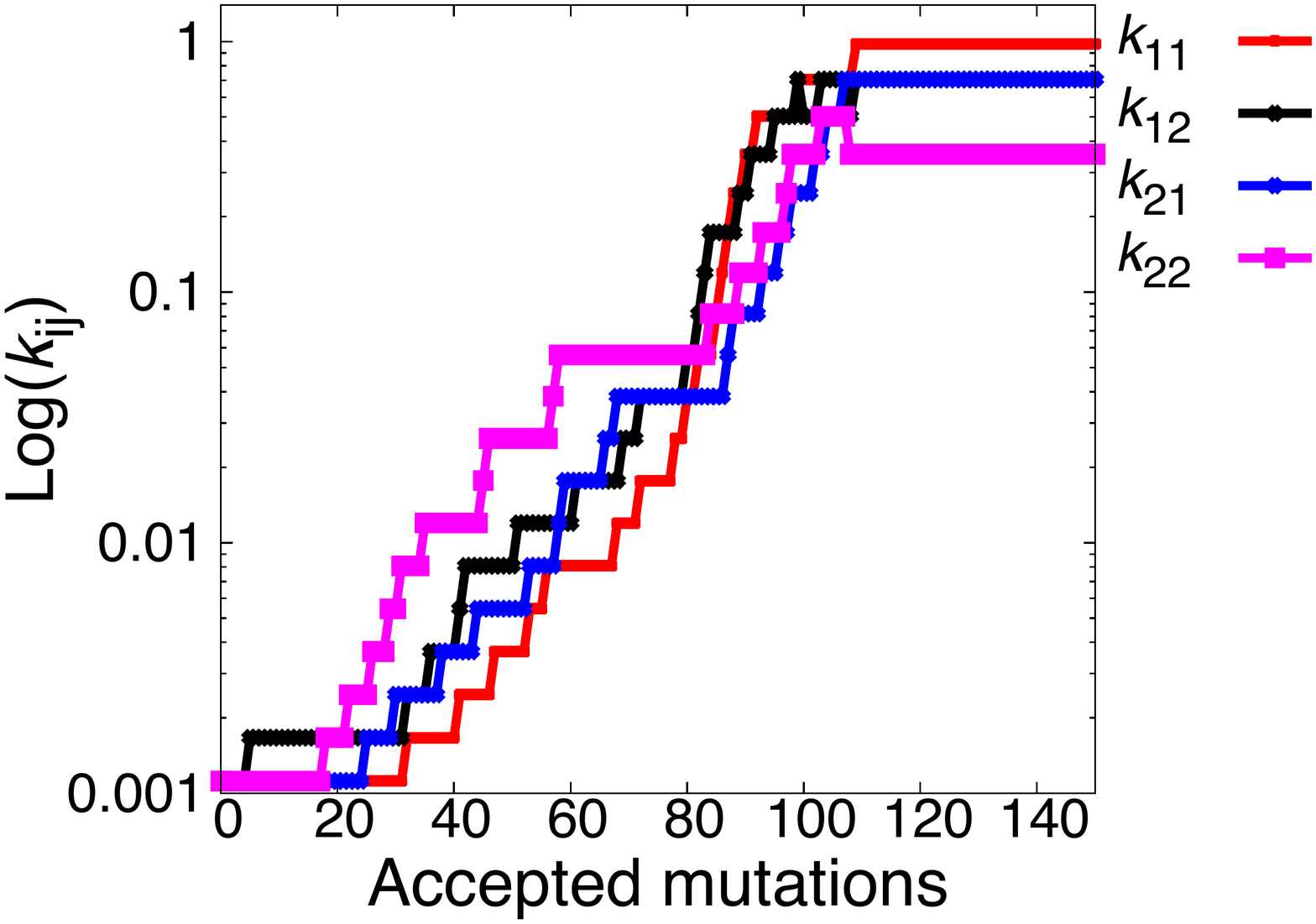}%
  \label{No_Bifurcation}
}

\subfloat[]{%
  \includegraphics[scale=0.255]{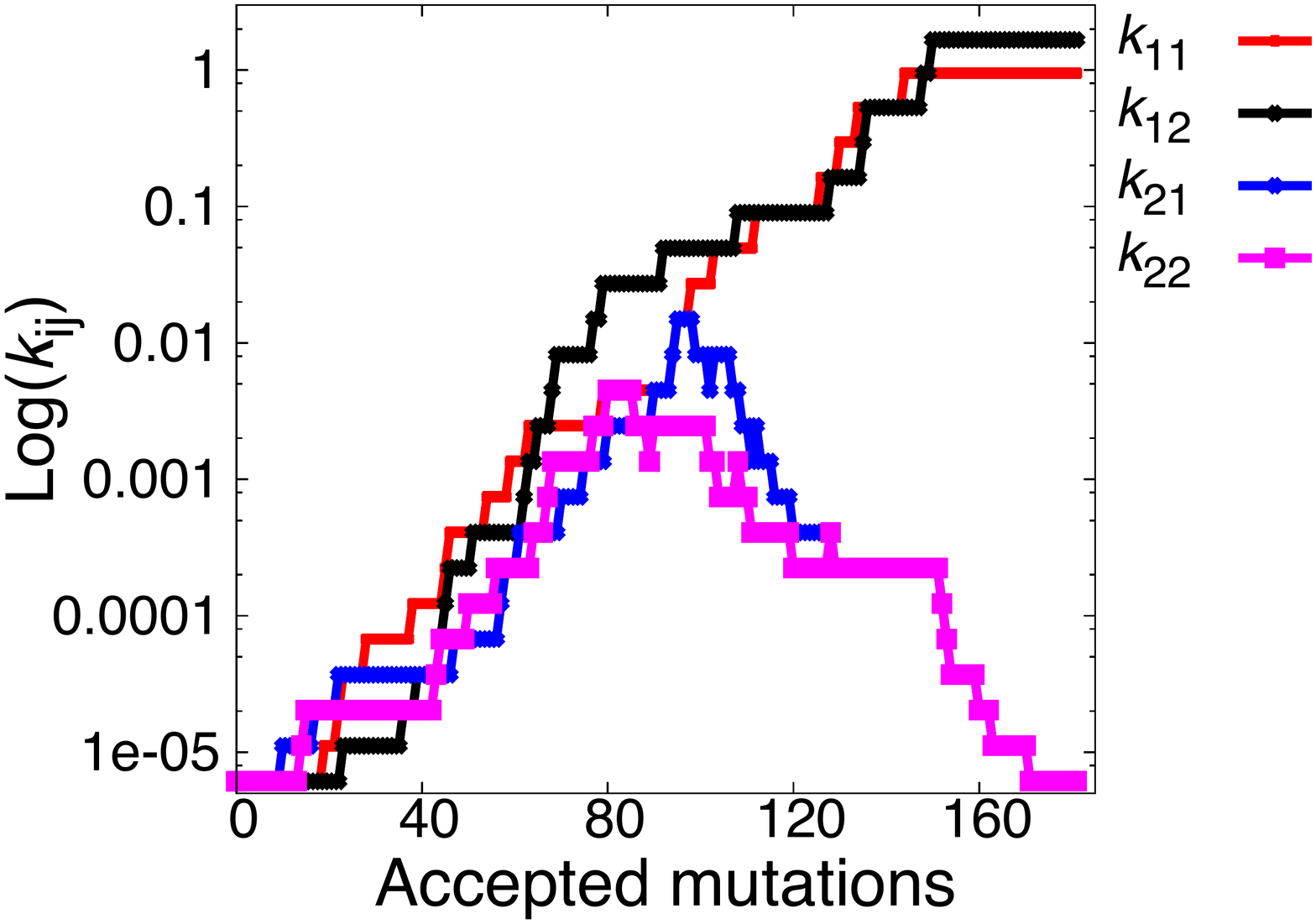}%
  \label{Bifurcation}
}

\subfloat[]{%
  \includegraphics[scale=0.255]{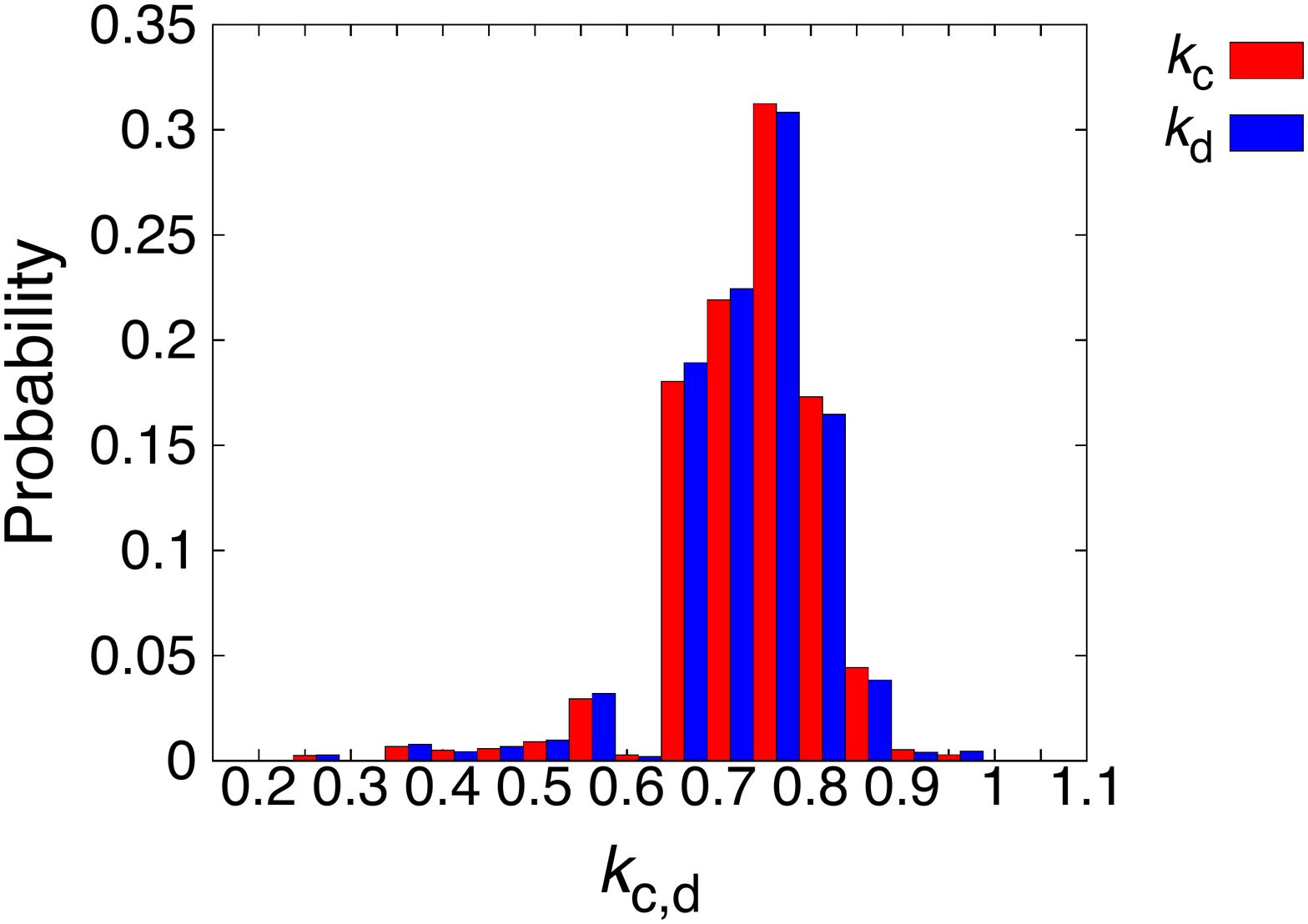}%
  \label{Histogram_Total_MI}

}
\caption{Evolution under $\mathrm{MI_{total}}$ and the probability distribution of rate constants $k_{ij}$. (a) Log($k_{\textrm{ij}}$) versus accepted mutations. (b) Bifurcations in pairs of rate constants for $\epsilon=0.6$. (c) Probability distribution of rate constants showing high degree of crosstalk for $\epsilon = 0.2$; constructed from 10,000 simulations. $k_\textrm{d}$ represents the direct rate constants, whereas $k_\textrm{c}$ represents the crosstalk rate constants. Other parameter values: $k_0 = 20$, $E_0 = 5$, $V=3$.}
\end{center}
\end{figure}

Surprisingly, evolving $\mathrm{MI_{total}}$ leaves the optimized network with a high degree of crosstalk, contrary to our expectations based on Fig. \ref{MI_2D_Density}. E.g. for the interaction parameter $\epsilon = 0.2$, if we start with low values of all $k_{ij}$, we typically find that all the rate constants increase simultaneously, as shown in Fig. \ref{No_Bifurcation}, implying high crosstalk. Strikingly, for larger $\epsilon$, the majority of runs exhibit bifurcations in rate constants, but still leave the optimized network with a high degree of crosstalk (see Fig. \ref{Bifurcation}). In a typical bifurcation, $k_{11}$ and $k_{12}$ might dominate while $k_{21}$ and $k_{22}$ are suppressed, whereas $k_{21}$ and $k_{22}$ might dominate in a different run. These bifurcations yield examples of signal ``fan-out" (single input, multiple outputs ) and signal ``fan-in" (multiple inputs, single output) \cite{papin}, found in biological systems \cite{chen_pleiotropic_1,Pagliari_pleiotropic_2,Granek_pleiotropic_3,gustin}. Fig. \ref{Histogram_Total_MI} shows a probability distribution of rate constants after rate constants have stopped changing under $\mathrm{MI_{total}}$ evolution; the peaks of the histogram occur at similarly high values of the crosstalk and direct rate constants, implying a high degree of crosstalk as an evolutionary outcome for $\mathrm{MI_{total}}$.

\begin{figure}[htp!]
\begin{center}
\subfloat[]{%
    \includegraphics[clip,width=0.8\columnwidth]{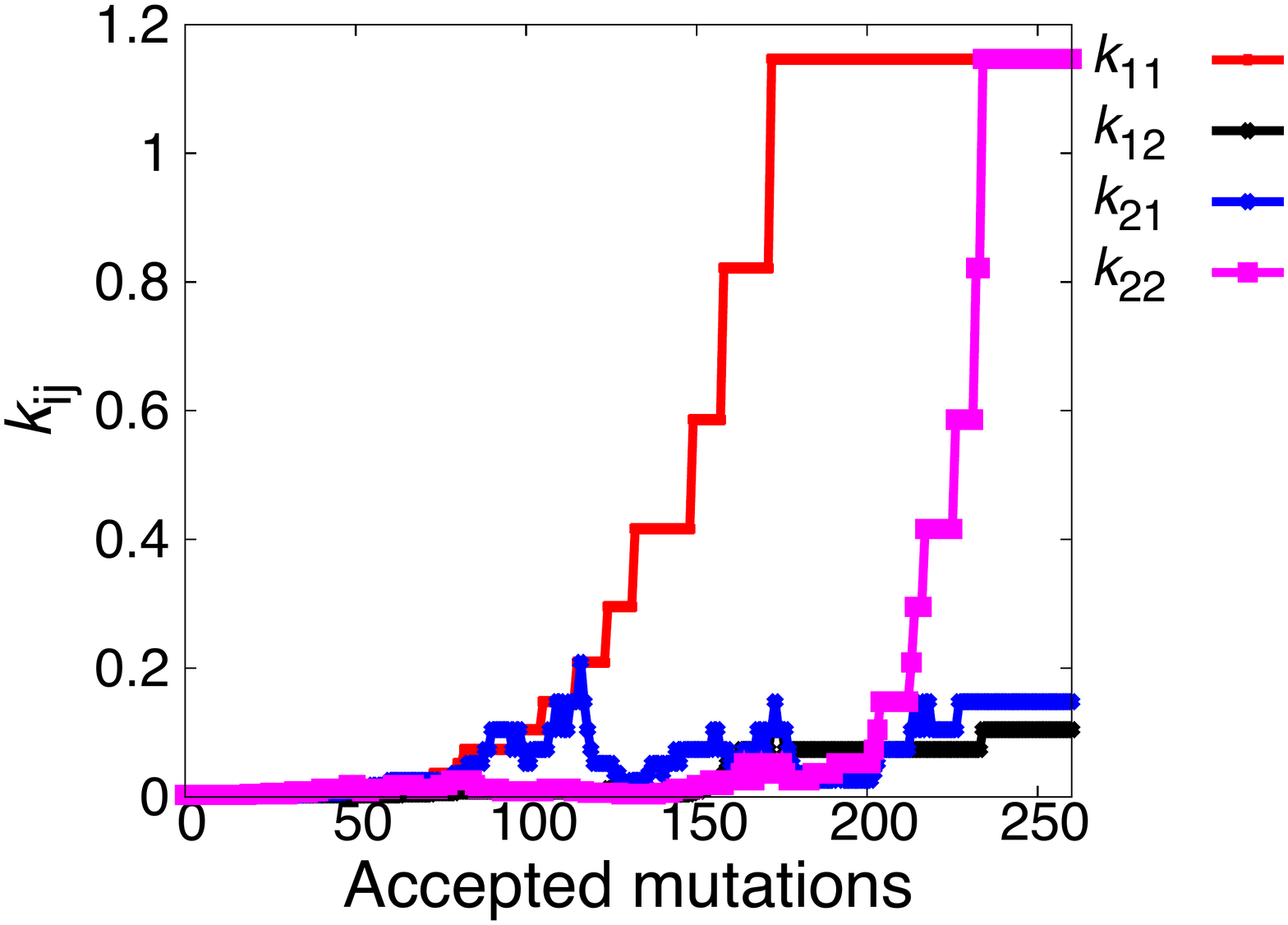}%
  \label{Suppressed_Crosstalk}
}

\subfloat[]{%
  \includegraphics[clip,width=0.8\columnwidth]{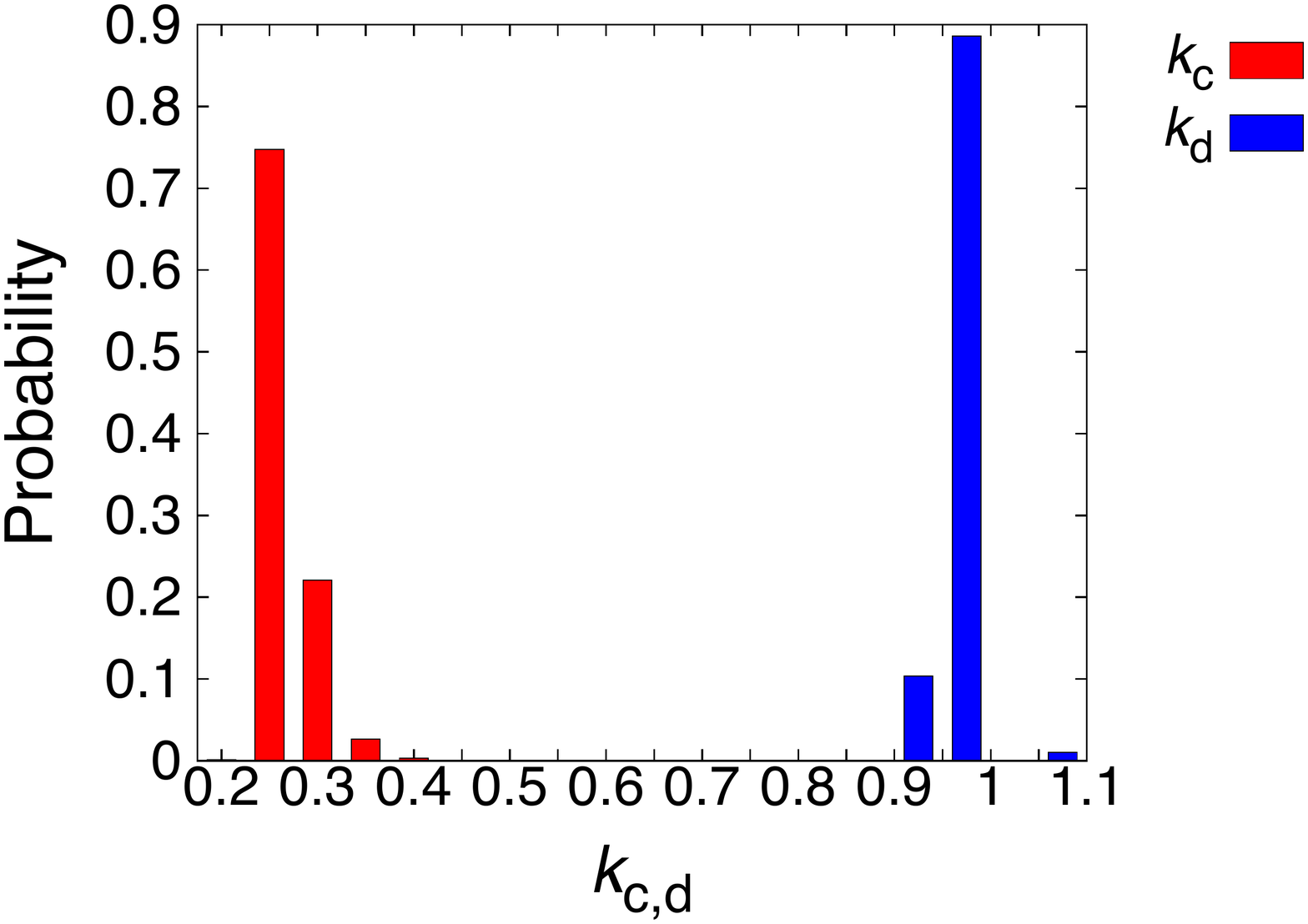}%
  \label{Histogram}

}
\caption{Evolution under $\mathrm{MI_{sum}}$ and the probability distribution of $k_{ij}$. (a) Rate constants versus accepted mutations. (b) Probability distribution of rate of constants showing suppression of crosstalk for $\epsilon = 0.2$; constructed from 10,000 simulations. $k_\textrm{d}$ represents the direct rate constants, whereas $k_\textrm{c}$ represents the crosstalk rate constants. }
\end{center}
\end{figure}

On the other hand, evolution under the fitness function $\mathrm{MI_{sum}}$ leads to low crosstalk and thus isolated pathways. Fig. \ref{Suppressed_Crosstalk} shows a typical run of greedy evolution under $\mathrm{MI_{sum}}$. Note that in this typical run, the direct rate constant values grow (e.g. $k_{11}$, $k_{22}$ $\sim 1$ in the evolved network, corresponding to the optimal values in the single-input single-output case, as in Fig. \ref{MI_Several_V_fixed_C}), whereas the crosstalk rate constants stay low (e.g. $k_{12}$, $k_{21}$ $\sim 0.1$). Fig. \ref{Histogram} shows a histogram exhibiting separation of crosstalk and direct rate constants, with high values of direct rate constants and low values of crosstalk rate constants. 

\captionsetup[subfigure]{singlelinecheck=false,justification=raggedright,position=top}

\begin{figure}
\includegraphics[scale=0.4]{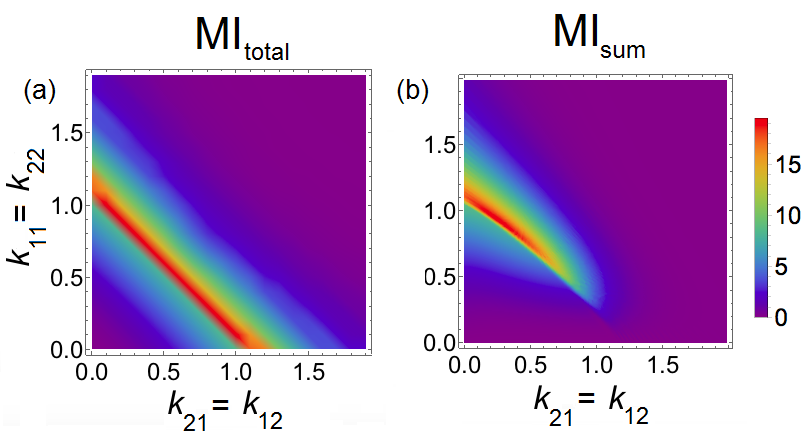}
\caption{The fitness landscapes plotted for rate constants $k_{21} = k_{12}$ and $k_{11} = k_{22}$, with $V=3$ for (a) $\mathrm{MI_{total}}$ and (b) $\mathrm{MI_{sum}}$. $\mathrm{MI_{total}}$ does not have a single global maximum associated with zero crosstalk whereas $\mathrm{MI_{sum}}$ does. 
}
\label{MI_4D_Landscapes}
\end{figure}

\begin{figure}
\begin{center}
\includegraphics[scale=0.35]{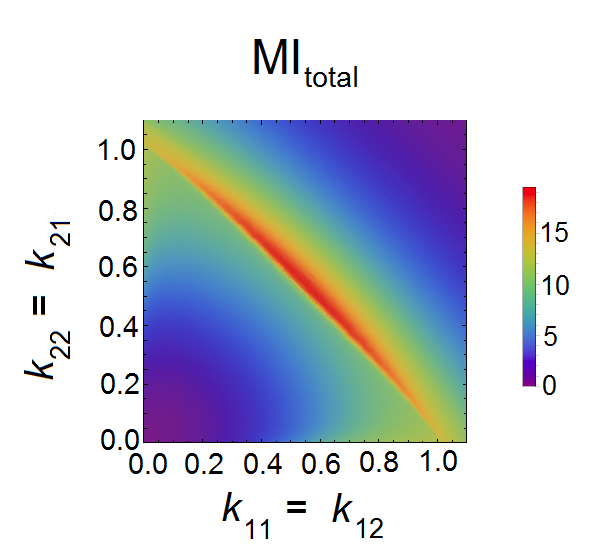}
\caption{ $\mathrm{MI_{total}}$ plotted as a function of rate constants $k_{11} = k_{12}$ and $k_{22} = k_{21}$ for $V=3$.}
\label{bifurcation_landscape}
\end{center}
\end{figure}

Our evolutionary approach has revealed essential differences between the two fitness functions. How can we understand the difference in evolutionary outcomes given that the maximum fitness depicted in Fig. \ref{MI_2D_Density} occurred at zero crosstalk for both fitness functions? Although the two landscapes appeared similar, it is important to recall that the phase space of the fitness landscapes is really four dimensional and Fig. \ref{MI_2D_Density} corresponds to a particular two-dimensional slice. We are then faced with the question of how to construct a lower dimensional slice of the fitness landscapes that could help us understand the difference in evolutionary outcomes. 
The crucial difference between evolutionary outcomes pertained to the typical ratio between direct and crosstalk rate constants; we therefore want to distinguish between the fitness dependence on the direct rate constants and crosstalk rate constants. Thus, we set $k_{11} = k_{22}$, corresponding to the direct rate constant, and  $k_{12} = k_{21}$, corresponding to the crosstalk rate constant, and construct a two-dimensional slice where one axis represents the direct rate constant and the other the crosstalk rate constant. As shown in Fig. \ref{MI_4D_Landscapes}, the resulting fitness landscapes reveal a striking difference between the two fitness functions. In particular, we note that while $\mathrm{MI_{sum}}$ is peaked at zero crosstalk (albeit with some spread to finite crosstalk), $\mathrm{MI_{total}}$ is optimal over an entire band corresponding to a range of direct and crosstalk rate constants. This observation helps us understand why the two fitness functions lead to very different evolutionary outcomes, and in particular, why $\mathrm{MI_{sum}}$ leads to low crosstalk while $\mathrm{MI_{total}}$ can result in a high degree of crosstalk. Lastly, to understand the bifurcations observed in rate constants for evolution under $\mathrm{MI_{total}}$ for larger values of $\epsilon$ (Fig. \ref{Bifurcation}), we construct another two-dimensional slice of the fitness landscape where we set $k_{11} = k_{12}$ and $k_{22} = k_{21}$ and plot the resulting $\mathrm{MI_{total}}$ in Fig. \ref{bifurcation_landscape}. We note that while the gradient of $\mathrm{MI_{total}}$ along the diagonal is positive, it can be smaller than the gradient along either axis so that $\mathrm{MI_{total}}$ could increase in the transverse direction away from the diagonal. For larger $\epsilon$, the change in the rate constants due a mutation could be larger, which increases the likelihood for the system to take a larger step away from the diagonal and to subsequently move towards either axis, leading to a bifurcation in the magnitudes of the rate constants.


\section{Discussion and Conclusion}
 
We have adapted a sequence based protein-protein interaction model to study the evolution of crosstalk in multiple-input, multiple-output signaling networks. Evolution is driven by random mutations in sequence space whereas selection occurs in the space of phenotypes. Interestingly, we have shown that two fitness functions, $\mathrm{MI_{total}}$ and $\mathrm{MI_{sum}}$, produce drastically different outcomes. $\mathrm{MI_{total}}$ represents the fitness for a system evolved to distribute the total information content of inputs throughout the signaling network whereas $\mathrm{MI_{sum}}$ represents fitness for a system where inputs are transmitted exclusively through their cognate signaling pathways. Using our evolutionary scheme we have shown that $\mathrm{MI_{total}}$ retains a high degree of crosstalk whereas $\mathrm{MI_{sum}}$ leads to insulated pathways with lowered crosstalk. In addition, we have seen how the evolutionary outcomes can be related to the fitness landscapes. In particular, we found that while $\mathrm{MI_{sum}}$ is optimized for zero crosstalk, $\mathrm{MI_{total}}$ is optimal over a range of crosstalk. Our results pertaining to dependence of $\mathrm{MI_{total}}$ on crosstalk are unique to biochemical channels where the strength of the noise depends on input; these results are different from Gaussian channels with constant additive noise where crosstalk always leads to reduction in total mutual information \cite{telatar}.

Our work focuses on stochasticity inherent to biochemical reactions (intrinsic noise) rather than variability in cellular states (extrinsic noise) \cite{Raser}. While generally both intrinsic and extrinsic noise degrade information transmitted through signaling networks, experiments show that signaling networks can mitigate, and potentially eliminate, extrinsic-noise-induced information loss \cite{jangir}. Furthermore, the impact of extrinsic noise decreases with increasing network complexity \cite{luca}, which justifies our focus on intrinsic noise (note however that owing to its simplicity, our framework can easily be generalized to incorporate extrinsic noise \cite{swain}). Our results are also robust to parameter choices. We varied our model parameters $k_0$, $\epsilon$, and $V$ such that the resulting rate constants $k_{ij}$ spanned three orders of magnitude and observed similar outcomes in our simulations. 

Our work shows that, depending on the choice of fitness function, evolution may or may not suppress crosstalk between signaling pathways. In biology we find systems displaying extensive crosstalk and also systems that exhibit high specificity. For example, signaling networks in eukaryotes display extensive crosstalk: during embryonic development of metazoans, complex and delicate interactions exist between the TGF-$\beta$/BMP, Wnt/Wg, Hedgehog (Hh), Notch, mitogen-activated protein kinase (MAPK), and other pathways \cite{attisano, sumi}. On the other hand, two component signaling networks, comprising the majority of prokaryotic signaling,  display very little crosstalk. The results in our paper imply that systems for which inputs have to be integrated in order to produce output, such as quorum sensing \cite{ned_quorum}, $\mathrm{MI_{total}}$ would be the appropriate fitness. In cases where distinct inputs require distinct responses from the system, we expect $\mathrm{MI_{sum}}$ to be the suitable quantity for fitness, in which case our results suggest  an evolutionary drive to eliminate crosstalk. An example for the latter is the high osmolarity and starvation response in yeast where the pathways respond to the appropriate environmental cues in very distinct and highly precise ways \cite{madhani,dohlman}, 

We expect our model to be broadly useful for exploring principles of protein network evolution. While simple and easy to implement, the model is biologically grounded in sequence-based evolution, and also physically grounded insofar as all proteins potentially interact with all others. While we have assumed completely uncorrelated input distributions for our system, it would be interesting to explore how correlated inputs might affect evolution of crosstalk. We have also focused on two-layer signaling processes, but these can readily be extended to include multilayer cascades \cite{kramer}. Future work will address the effects of adding feedback, a higher number of pathways, and proteins such as histidine kinases that act both as activators and deactivators \cite{wolanin}.

This work was supported in part by the National Science Foundation, Grant PHY-1305525, and the National Institutes of Health,  Grant R01 GM082938.


\clearpage
\pagebreak
\widetext

\setcounter{equation}{0}
\setcounter{figure}{0}
\setcounter{table}{0}
\setcounter{page}{1}

\setcounter{section}{0}

\hyphenation{ALPGEN}
\hyphenation{EVTGEN}
\hyphenation{PYTHIA}

\begin{center}
\textbf{\large Supplementary Material: Modeling Evolution of Crosstalk in Noisy Signal Transduction Networks}
\end{center}
\author{Ammar Tareen}
\affiliation{Clark University, Department of Physics, Worcester, MA 01610}
\author{Ned S. Wingreen}
\affiliation{Lewis-Sigler Institute for Integrative Genomics, Carl Icahn Laboratory, Washington Road, Princeton, NJ 08544}
\author{Ranjan Mukhopadhyay}
\affiliation{Clark University, Department of Physics, Worcester, MA 01610}

\maketitle



\renewcommand\theequation{S\arabic{equation}}
\renewcommand{\thefigure}{S\arabic{figure}}

\section{Langevin and Fokker-Planck (FP) Equations}

\begin{figure}[H]
\begin{center}
\includegraphics[width = 7 cm, height =  0.8 cm]{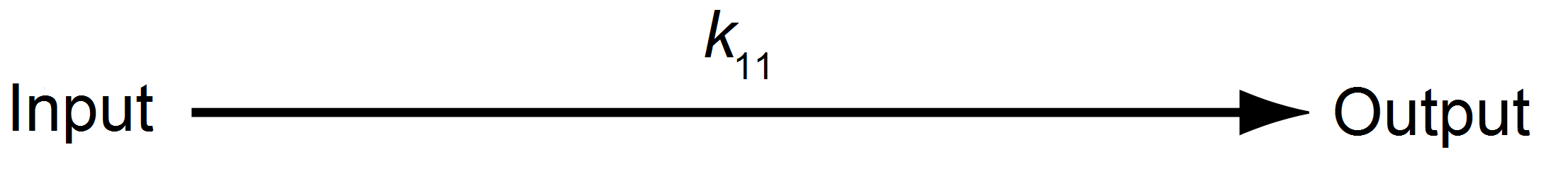}
\caption{Single signaling pathway.}
\label{1D_Channel}
\end{center}
\end{figure}
Consider a one-input, one-output system (Fig. \ref{1D_Channel}), and let $n(t)$ represent the number of activated output proteins, $n_0$ represent the total number of output proteins, and $V$ be the volume of the system. We can write a Langevin equation for information transmission along a single signaling pathway in terms of $n(t)$ as follows
\begin{equation}
\frac{dn}{dt} = k_{11} I(n_0-n)-\alpha n +\sqrt{k_{11}I(n_0-n) +\alpha n}\xi(t).
\end{equation}
If we now use the relations $n_0=O_\mathrm{tot} V$ and $n=O^* V$ to change the equation for output protein number into an equation for concentration, we obtain the Langevin equation from the main text 
 \begin{equation}
\frac{dO^*}{dt} = k_{11} IO-\alpha O^* +\sqrt{\frac{k_{11}I O +\alpha O^*}{V}}\xi(t),
\label{Langevin_ST}
\end{equation}
where $O_\mathrm{tot} = O + O^*$ is constant. The Langevin equation can be written as a deterministic part $A$ and a stochastic part $B$:
\begin{equation}
\frac{dO^*}{dt} = A(O^*,t)+B(O^*,t)\xi(t),
\end{equation}
where functions $A$ and $B$ are as follows,
\begin{equation}
A(O^*,t) = k_{11}IO-\alpha O^*,
\end{equation}
\begin{equation}
B(O^*,t) = \sqrt{\frac{k_{11}IO +\alpha O^*}{V}}.
\end{equation}
From the Langevin equation one can arrive at an FP equation that governs the time evolution of conditional probability $P(O^*|I, k_{11})$. However, it is important to note that a Langevin equation with a concentration dependent noise coefficient (function $B$) may lead to two different forms of the Fokker-Planck equation depending on the choice of time used in a Wiener process. A choice of midpoint time yields the Stratonovich calculus, which results in the Stratonovich interpretation of the Fokker-Planck equation \cite{risken,vanKampen}:

\begin{equation}
\frac{\partial P}{\partial t} = -\frac{\partial }{\partial O^*} \Big \{  \Big[ A(O^*,t) + B(O^*,t) \frac{\partial B(O^*,t)}{\partial O^*} \Big] P \Big\}+
 \frac{1}{2}\frac{\partial^2 }{\partial {O^*}^2} \Big\{ \Big[ B^2(O^*,t) \Big] P \Big\}.
\label{FPS}
\end{equation}
A choice of time at the beginning of the integration interval yields the It\^o form of the Fokker-Planck equation:
\begin{equation}
\frac{\partial P}{\partial t} = -\frac{\partial }{\partial O^*} \Big \{  \Big[ A(O^*,t) \Big] P \Big\}+
 \frac{1}{2} \frac{\partial^2 }{\partial {O^*}^2} \Big\{ \Big[ B^2(O^*,t) \Big] P \Big\}.
\label{FPI}
\end{equation}
Note that in the Stratonovich formulation, Eq. \ref{FPS}, there is an extra term of the form $ B(O^*,t) \frac{\partial B(O^*,t)}{\partial O^*}$. Whether such a term should be present in the Fokker-Planck equation can be resolved by writing down the master equation governing the physical process under question and using the Kramers-Moyal expansion to arrive at the Fokker-Planck equation. In our case, the resulting Fokker-Planck is of the form of Eq. \ref{FPI}, without the additional term, thus justifying the It\^o formulation \cite{Garcia-Palacios}. These considerations generalize easily to multivariate systems, for example, see Chapter 5 in \cite{Garcia-Palacios}.

\section{Solution of the Fokker-Planck Equation}

As we have seen, the Fokker-Planck equation can be written in the form:
\begin{equation}
\frac{\partial P}{\partial t} = -\frac{\partial }{\partial O^*} \Big \{  A(O^*,t) P \Big\}+
 \frac{1}{2}\frac{\partial^2 }{\partial {O^*}^2} \Big\{  B^2(O^*,t)  P \Big\}.
\label{FP1}
\end{equation}
We are interested in finding the steady-state probability, which means the left hand side of Eq. \ref{FP1} is zero. Extracting one derivative with respect to $O^*$, the equation becomes

\begin{equation}
0 = -\frac{\partial }{\partial O^*} \Big[ \Big \{   A(O^*,t) P \Big\}-
\frac{1}{2} \frac{\partial}{\partial O^*} \Big\{ \Big[ B^2(O^*,t) \Big] P \Big\} \Big],
\end{equation}
which means that the expression inside the outer square brackets must equal a constant, and we can reduce the second-order differential to a first-order differential equation:

\begin{equation}
C_1 = \Big \{   A(O^*,t)  P \Big\}-
\frac{1}{2} \frac{\partial}{\partial O^*} \Big\{ \Big[ B^2(O^*,t) \Big] P \Big\}.
 \label{FP_1st_Order}
\end{equation}
Here $C_1$ is an arbitrary constant. Before we proceed, note is that the Fokker-Planck equation has the form

\begin{equation}
\frac{\partial P}{\partial t} = -\frac{\partial }{\partial O^*} {AP} +  \frac{1}{2}\frac{\partial^2 }{\partial {O^*}^2}{B^2 P}, 
\end{equation}
which can be re-written as  

\begin{equation}
\frac{\partial P}{\partial t} = -\frac{\partial }{\partial O^*}( \underbrace{{AP} -  \frac{1}{2}\frac{\partial }{\partial O^*}{B^2 P}}_\text{J} ).
\end{equation}
Denoting the term in the parenthesis as $J$, we obtain the continuity equation representing the conservation of probability

\begin{equation}
\frac{\partial P}{\partial t} + \nabla{J} = 0,
\end{equation}
where $\nabla$ corresponds to $\frac{\partial }{\partial O^*}$. $J$ is the probability current, which must vanish at the boundaries of $O^*$ and therefore will provide the boundary conditions. Eq. \ref{FP_1st_Order} can now be written as 
\begin{equation}
C_1 = AP - BB'P - \frac{1}{2}B^2P',
\end{equation}
where the prime denotes a derivative with respect to $O^*$. Rewriting, 
\begin{equation}
C_1 = (\underbrace{A - BB'}_\text{$a$})P - (\underbrace{\frac{1}{2}B^2}_\text{$b$})P',
\end{equation}

\begin{equation}
C_1 = a(O^*)P - b(O^*)P'.
\end{equation}
We divide through by $b(O^*)$ to obtain (not writing the $O^*$ dependence explicitly, for convenience)

\begin{equation}
\underbrace{\frac{C_1}{b}}_\text{$G$} = \underbrace{\frac{a}{b}}_\text{$F$}P - P',
\end{equation}
we then have,

\begin{equation}
P'(O^*)-F(O^*)P(O^*) = -G(O^*). 
\label{IntFactor}
\end{equation}
Now assume existence of an integrating factor $\mu(O^*)$ with the property that $\mu F = -\mu'$. We multiply Eq. \ref{IntFactor} by $\mu$ to obtain

\begin{equation}
\mu(O^*)P'(O^*)-\mu(O^*)F(O^*)P(O^*) = -\mu(O^*)G(O^*). 
\end{equation}
Using the defining property of $\mu$, we have

\begin{equation}
\mu(O^*)P'(O^*)+\mu'(O^*)P(O^*) = -\mu(O^*)G(O^*). 
\end{equation}
Now we rewrite this as 

\begin{equation}
[\mu(O^*)P(O^*)]' = -\mu(O^*)G(O^*). 
\end{equation}
Integrating both sides,

\begin{equation}
\int[\mu(O^*)P(O^*)]'dO^* = -\int\mu(O^*)G(O^*)dO^*,
\end{equation}
we have

\begin{equation}
\mu(O^*)P(O^*)+C_2 = -\int\mu(O^*)G(O^*).
\end{equation}
Now

\begin{equation}
P(O^*) = \frac{-(\int\mu(O^*)G(O^*) +C_2)}{\mu(O^*)}.
\label{general_sol_mu}
\end{equation}
To determine $\mu$, we go back to the definition of $\mu$ and write

\begin{equation}
-\frac{\mu'(O^*)}{\mu(O^*)} = -\ln{(\mu(O^*))}' =  F(O^*).
\end{equation}
Integrating both sides again and multiplying by $-1$, we have

\begin{equation}
\ln{(\mu(O^*))} - C_3 =  -\int F(O^*)dO^*.
\end{equation}

\begin{equation}
\ln{(\mu(O^*))} =  -\int F(O^*)dO^*+C_3.
\end{equation}
Exponentiating 

\begin{equation}
\mu(O^*) =  e^{-\int F(O^*)dO^*}e^{C_3},
\end{equation}
renaming the constant 

\begin{equation}
\mu(O^*) =  C_3 e^{-\int F(O^*)dO^*},
\label{mu}
\end{equation}
and inserting \ref{mu} into \ref{general_sol_mu}, we obtain

\begin{equation}
P(O^*) = \frac{-(\int C_3 e^{-\int F(O^*)dO^*}G(O^*) +C_2)}{C_3 e^{-\int F(O^*)dO^*}}.
\label{general_sol_extra_const}
\end{equation}
Extracting out a factor of $C_3$, we have

\begin{equation}
P(O^*) = \frac{-(\int  e^{-\int F(O^*)dO^*}G(O^*) +\frac{C_2}{C_3})}{e^{-\int F(O^*)dO^*}},
\label{general_sol_extra_const}
\end{equation}
and we denote the ratio $\frac{C_2}{C_3}$ as $C_4$ to write

\begin{equation}
P(O^*) = \frac{-(\int  e^{-\int F(O^*)dO^*}G(O^*) +C_4)}{e^{-\int F(O^*)dO^*}},
\label{general_sol_FG}
\end{equation}
which is the solution of the Fokker-Planck equation for the one-input, one-output system for arbitrary boundary conditions. 

\subsection{FP Solution at $J=0$}

We have noted that the probability current $J$ is a constant in steady state. In terms of boundary conditions that $J$ needs to satisfy, we note that $J$ is a function of $O^*$ and $O^*$ can vary in the range [0-1]. Since the system does not have any sources or sinks at either boundary of $O^*$, we expect $J$ to be zero at both boundaries (this is akin to reflective boundary conditions). Because $J$ needs to be constant, the only possible solution is that it is 0 everywhere. However, since our state space of activated protein concentration is finite, the probability need not be zero at the boundary, only the probability current needs to be zero. \\

Continuing from Eq. \ref{FP_1st_Order} in the previous section, the Fokker-Planck equation now becomes

\begin{equation}
\Big \{  \Big[ A(O^*,t) \Big] P \Big\}-
\frac{1}{2}\frac{\partial}{\partial O^*} \Big\{ \Big[ B^2(O^*,t) \Big] P \Big\} = 0.
\end{equation} 
Simplifying notation, and following similar algebra to the previous section, we have:

\begin{equation}
AP - BB'P - \frac{1}{2}B^2P' = 0,
\end{equation}
which is readily solved by 

\begin{equation}
P(O^*|k_{11}) = C_1 e^{\int (\frac{2(A-BB')}{B^2})}.
\end{equation}
For convenience, we can denote $C_1$ as $N$ (normalization constant). To determine the value of the normalization constant, we integrate over the entire concentration interval $[0,1]$, and set the result to 1. So the output probability distribution, after substituting in the forms of $A$ and $B$ and after some algebra, can be written as


\begin{equation}
P(O^* | I,k_{11})  = N {e^{\frac{2 V O^* (I^2 k_{11}^2-\alpha^2)}{(\alpha-Ik_{11})^2}}} {\Big[1+\frac{(\alpha-I k_{11})O^*)}{I k_{11} O_{\textrm{tot}}}\Big]}^{{\frac{4 I k_{11} O_{\textrm{tot}} \alpha V}{(\alpha-I k_{11})^2}}-1}.
\label{P_y_given_x}
\end{equation}
Obtaining the solution of the multivariate system is now quite straightforward and follows almost exactly the same procedure. 

\subsection{FP Solution at $I = \frac{\alpha}{k_{11}}$}

Exactly at the point $I = \frac{\alpha}{k_{11}}$, the function $B(O^*,t)$ is independent of $O^*$, and the conditional probability at that point simplifies to 

\begin{equation}
P(O^* | I = \frac{\alpha}{k_{11}}, k_{11}) =N e^{2 V O^* -\frac{2 V {O^*}^2}{C}}.
\end{equation}

\subsection{FP solution near $I = \frac{\alpha}{k_{11}}$}

One might worry that the expression for probability in Eq. \ref{P_y_given_x} diverges as $\alpha \rightarrow I k_{11}$. This section shows that by setting $\delta = \alpha-Ik_{11}$ and Taylor expanding Eq. \ref{P_y_given_x} around $\delta=0$, the terms that appear to diverge actually cancel out leaving a finite-valued function. We thus rewrite Eq. \ref{P_y_given_x} as

\begin{equation}
P(O^* | I,k_{11})  = N e^{\overbrace{\frac{-2 V O^* (I k_{11}+\alpha)}{(\alpha-Ik_{11})}}^\text{{\large $\tilde{a}$}}}e^{\overbrace{-\log[1+\frac{(\alpha-I k_{11}) O^*)}{I k_{11} O_{\textrm{tot}}}]}^\text{{\large $\tilde{b}$}}}e^{\overbrace{\frac{4 I k_{11} O_{\textrm{tot}} \alpha V}{(\alpha-I k_{11})^2}\log[1+\frac{(\alpha-I k_{11})O^*)}{I k_{11} O_{\textrm{tot}}}]}^\text{{\large $\tilde{c}$}}}.
\label{P_y_given_xExp}
\end{equation}
For notational convenience, we define terms $\gamma \equiv \frac{O^*}{Ik_{11}O_{\textrm{tot}}}$, and $\beta \equiv  4 I k_{11} O_{\textrm{tot}} \alpha V$. Note that the expression ``$\tilde{b}$" in Eq. \ref{P_y_given_xExp} does not include any terms that might potentially diverge, so we focus on terms ``$\tilde{a}$" and ``$\tilde{c}$". Rearranging and rewriting, we have  

\begin{equation}
P(O^* | I,k_{11})  = N e^{{\frac{- (\beta \gamma)(2\alpha-\delta)}{(2 \alpha \delta)}}+{{\frac{\beta}{(\delta)^2}\log(1+\gamma \delta)}}}e^{{-\log(1+\gamma \delta)}},
\label{P_y_given_xExp2}
\end{equation}
which is equivalent to ${N \exp({\tilde{a}+\tilde{c}}) \exp{\tilde{(b)}}}$. It might appear that $\tilde{a}+\tilde{c}$ diverges as $\delta$ approaches zero. To check this, we Taylor expand $\tilde{a}+\tilde{c}$ in powers of $\delta$ around $\delta =0$, using the expansion for log$(1+\gamma \delta$). We find 

\begin{equation}
\tilde{a}+\tilde{c}  \approx {-\frac{\beta \gamma (2 \alpha - \delta)}{2 \alpha \delta}+\frac{\beta \gamma}{\delta}}+\mathcal{O}(\delta).
\label{P_y_given_xExp3}
\end{equation}
We can then see that the divergent terms cancel:
\begin{equation}
\tilde{a}+\tilde{c}  \approx  {-\frac{\beta \gamma}{2\alpha}+\cancel{\frac{\beta \gamma}{\delta}}-\cancel{\frac{\beta \gamma}{\delta}}}+\mathcal{O}(\delta).
\label{P_y_given_xExp4}
\end{equation}

\section{Chemical Rate Equations for Signaling Systems}

\begin{figure}[H]
\begin{center}
\includegraphics[scale = 0.5]{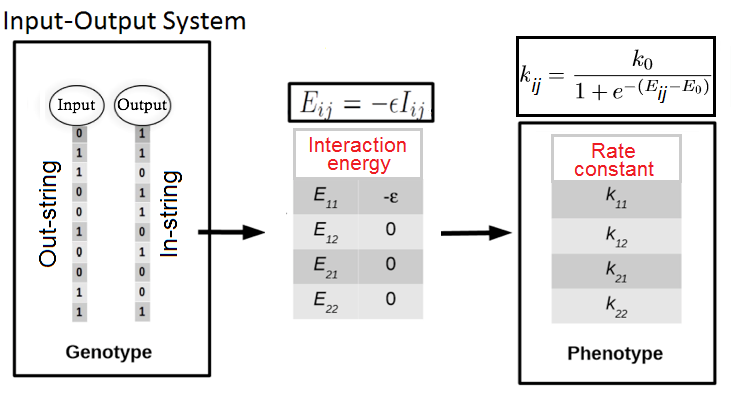}
\caption{Genotype to phenotype map depicting how 0's and 1's in the binary sequences interact to determine binding energies $E_{ij}$. A 1-1 interaction produces an interaction energy equal to $\epsilon$. All other interactions contribute zero interaction energy. }
\label{geno_pheno}
\end{center}
\end{figure}

The chemical reactions governing our single-input, single-output system are written as
\begin{equation}
I^* + O \overset{k_f}{\underset{k_{r}}{\rightleftharpoons}} I^*O  \overset{r}{\rightarrow}I^* +O^*, \hspace{0.5 cm} O^* \overset{\alpha}{\rightarrow} O,
\end{equation}
where $I^* O$ represents the intermediate complex. The rate of change of concentration of the active fraction of output is 
\begin{equation}
\frac{d [O^*]}{dt} = r[I^* O] - \alpha[O^*],
\label{qs2}
\end{equation}
where, for example, $[O^*]$ represents the concentration of $O^*$. Under the assumption that the intermediate complex concentration is at steady state (quasi-static approximation), we obtain
\begin{equation}
0 = \frac{d [I^*O]}{dt} = k_{f} [I^*][O]-(k_{r}+r)[I^* O],
\end{equation}
which implies
\begin{equation}
[I^* O] = \frac{k_{f} }{k_{r}+r} [I^*][O]. 
\label{qs1}
\end{equation}
Substituting Eq. \ref{qs1} into Eq. \ref{qs2}, we obtain
\begin{equation}
\frac{d [O^*]}{dt} = \frac{k_f }{1 +\frac{k_r}{r}} [I^*][O] - \alpha[O^*]  = k_{11} [I^*][O] - \alpha[O^*],
\end{equation}
where $k_{11}= \frac{k_f }{1 +\frac{k_r}{r}} $. For convenience, in the main text, we suppress the square brackets and denote the concentrations just as $O^*$ and $I^*$. Following the supplementary material in \cite{zulfikar1}, we assume an Arrhenius-type form for the rate $k_r = k_{r,0} e^{-\beta E_\textrm{{int}}}$, where $k_{r,0}$ is a constant and $\beta = \frac{1}{k_B T}$, which implies
\begin{equation}
k_{11} = \frac{k_{f}}{\frac{k_{{r,0}}}{r} e^{-\beta E_\textrm{{11}}}+1}, 
\end{equation}
where $E_\textrm{11}$ is the interaction energy. We can also write this equation in the form
\begin{equation}
k_\textrm{11} = \frac{k_0}{1+e^{-(E_\textrm{11}-E_0)}},
\end{equation}
where $k_0$ and $E_0$ are constants, and the energy $E_{11}$ is expressed in units of $k_B T$. These equations generalize readily to a system with multiple channels. As mentioned in the main text, in our evolutionary scheme, the interaction energy $E_{ij}$ is determined by the interaction between the in-string of output protein (denoted by index j) and the out-string of input protein (denoted by index i). Fig. \ref{geno_pheno} depicts an image of how the binary sequences (genotype) of 1s and 0s in our system interact to give rise to binding energies $E_{ij}$ that determine the values of the rate constants $k_{ij}$ (phenotype). 

\section{Initial Conditions}
 
\begin{figure}[H]
\begin{center}
\begin{floatrow}
\ffigbox{%
  \includegraphics[scale=0.23]{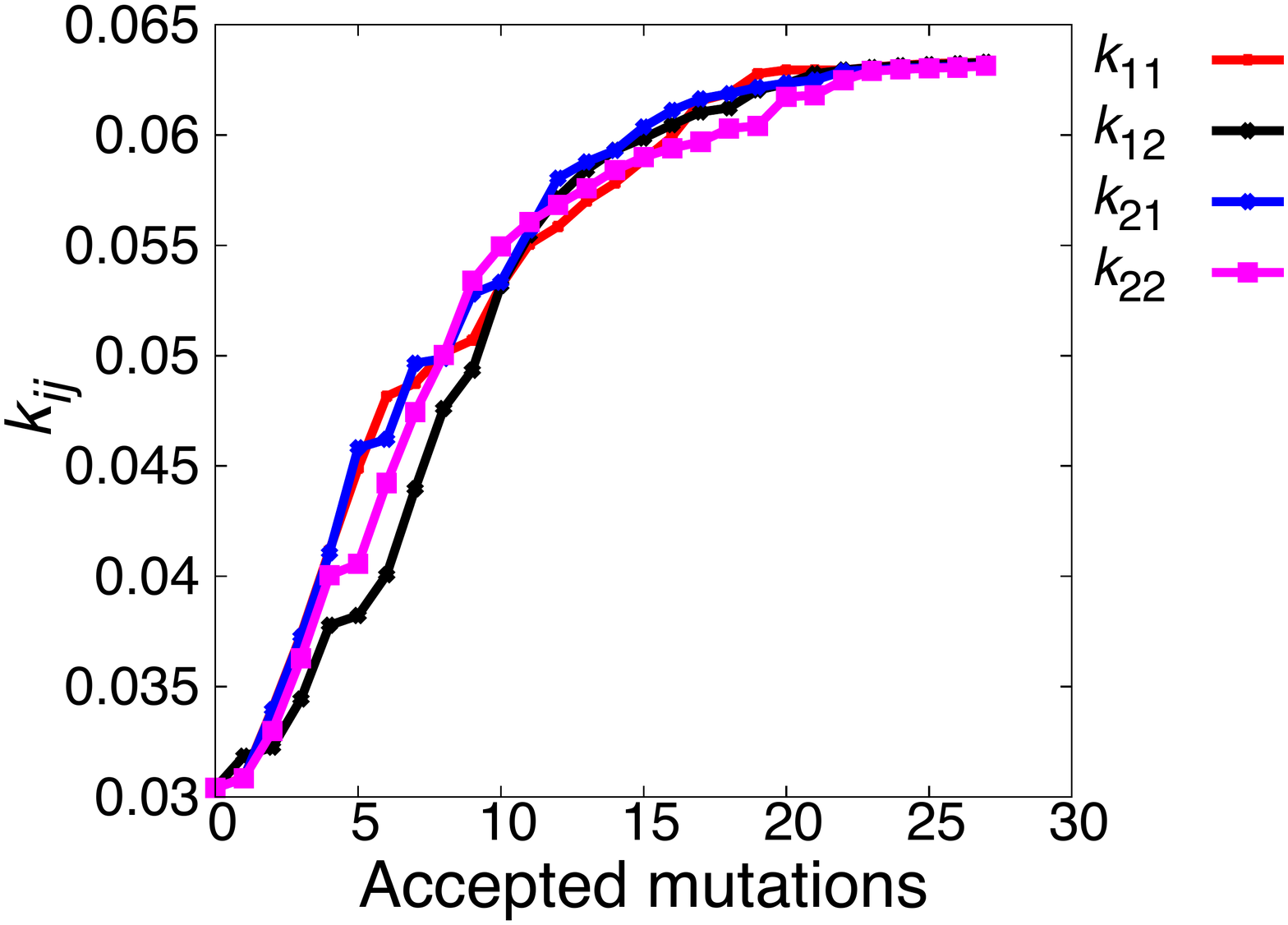}
}{%
  \caption{Rate constants versus accepted mutations. Duplicated initial strings.}%
  \label{Strict_IC_1_Duplicated}
}
\ffigbox{%
  \includegraphics[scale=0.24]{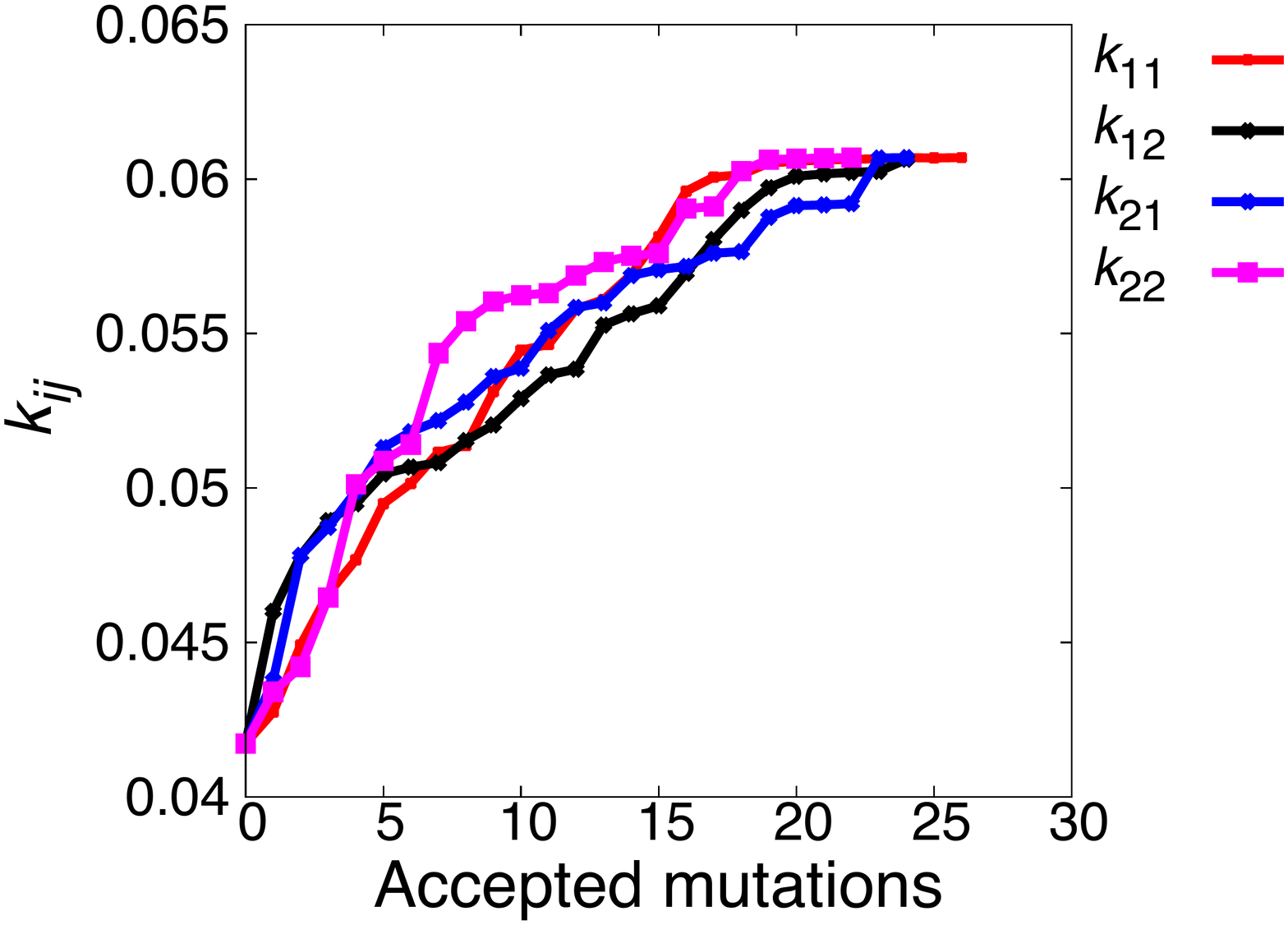}
}{%
  \caption{Rate constants versus accepted mutations. arbitrary initial strings.}
  \label{Strict_IC_2_arbitrary}
}
\end{floatrow}
\end{center}
\end{figure}

 From the main text, we know that the final values of the rate constants do not depend critically on our choice of initial strings. Fig. \ref{Strict_IC_1_Duplicated} and Fig. \ref{Strict_IC_2_arbitrary} show greedy evolution under $\mathrm{MI_{total}}$ with duplicated initial and completely arbitrary initial strings as initial conditions, respectively. In our simulations for one channel, initial strings were generated randomly. For two channels, one in-string and one out-string were generated randomly, and then both strings were duplicated. For reference, some initial strings used in simulations are listed below: \\

	\noindent $\mathrm{In}_1$ \hspace{1.5 mm} = (0,0,0,0,0,0,0,0,0,0,0,0,0,0,0,0,0,0,0,0,0,0,0,0,0) \\
	$\mathrm{In}_2$ \hspace{1.5 mm}  = (0,0,0,0,0,0,0,0,0,0,0,0,0,0,0,0,0,0,0,0,0,0,0,0,0) \\
	$\mathrm{Out}_1$  = (0,0,0,0,0,0,0,0,0,0,0,0,0,0,0,0,0,0,0,0,0,0,0,0,0) \\
	$\mathrm{Out}_2$  = (0,0,0,0,0,0,0,0,0,0,0,0,0,0,0,0,0,0,0,0,0,0,0,0,0) \\
		
	\noindent $\mathrm{In}_1$ \hspace{1.5 mm} =   (0,0,1,0,0,1,0,0,1,0,0,0,0,1,0,0,0,1,0,0,0,0,0,0,0) \\
	$\mathrm{In}_2$ \hspace{1.5 mm} =  (0,0,1,0,0,1,0,0,1,0,0,0,0,1,0,0,0,1,0,0,0,0,0,0,0) \\
	$\mathrm{Out}_1$ = (0,0,1,0,0,1,0,0,1,0,0,0,0,1,0,0,0,1,0,0,0,0,0,0,0) \\
	$\mathrm{Out}_2$ = (0,0,1,0,0,1,0,0,1,0,0,0,0,1,0,0,0,1,0,0,0,0,0,0,0) \\
	
	\noindent	$\mathrm{In}_1$ \hspace{1.5 mm} =  (0,0,1,1,1,1,0,1,0,1,1,1,1,1,1,0,1,0,1,1,1,1,1,0,1) \\
	$\mathrm{In}_2$ \hspace{1.5 mm} =  (0,0,1,1,1,1,0,1,0,1,1,1,1,1,1,0,1,0,1,1,1,1,1,0,1) \\
	$\mathrm{Out}_1$ = (1,1,1,1,0,1,0,1,0,0,1,1,1,1,1,0,1,1,0,1,1,1,1,0,1) \\
	$\mathrm{Out}_2$ = (1,1,1,1,0,1,0,1,0,0,1,1,1,1,1,0,1,1,0,1,1,1,1,0,1) \\

\section{Metropolis-Hastings Monte Carlo}

\begin{figure}[H]
\begin{center}
\includegraphics[scale=0.25]{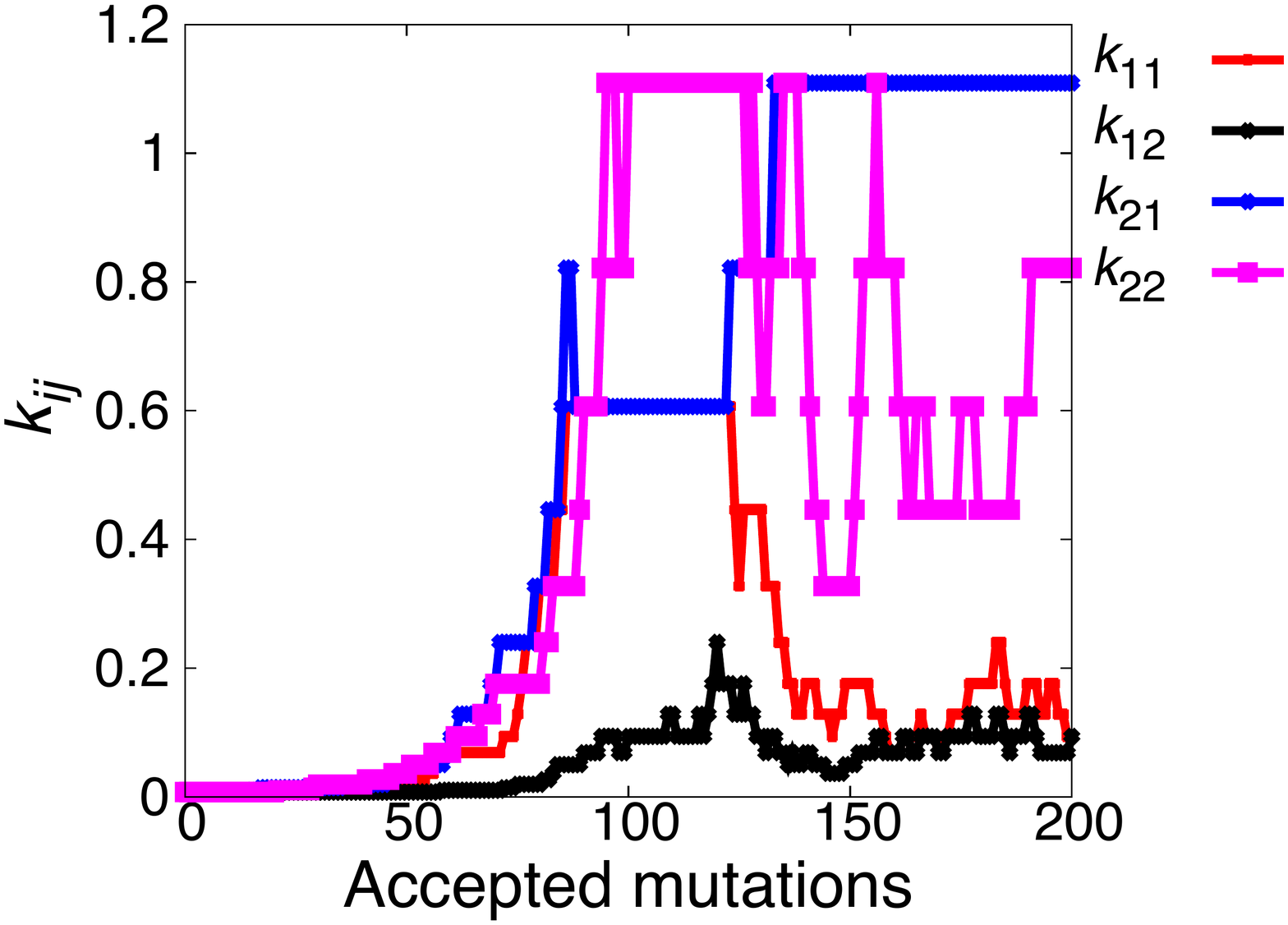}
\caption{A sample Monte Carlo run shown where crosstalk is not suppressed under $\mathrm{MI_{total}}$, using $\tilde{\beta}=0.005$.}
\label{Monte_Carlo}
\end{center}
\end{figure}

As a test of whether the failure of $\mathrm{MI_{total}}$ to suppress crosstalk was an artifact of our greedy evolutionary algorithm, we also employed a Monte-Carlo algorithm as our evolutionary scheme. Fig. \ref{Monte_Carlo} shows a sample run where a Metropolis-Hastings algorithm was implemented for evolution under $\mathrm{MI_{total}}$. The algorithm accepts all positive changes in fitness and accepts negative changes in fitness with a Boltzmann-like probability $e^{-\tilde{\beta} \mathrm{MI_{total}}}$, and otherwise rejects the changes. $\tilde{\beta}$ is a tunable parameter akin to the selection pressure ($\tilde{\beta}$ is equivalent to temperature in equilibrium statistical mechanics systems). We used a range of $\tilde{\beta}$ between (0-10). 

\section{Robustness of Fitness Landscapes to Strength of Noise}

\begin{figure}[H]
\begin{center}
\subfloat[]{%
    \includegraphics[clip,width=0.3\columnwidth]{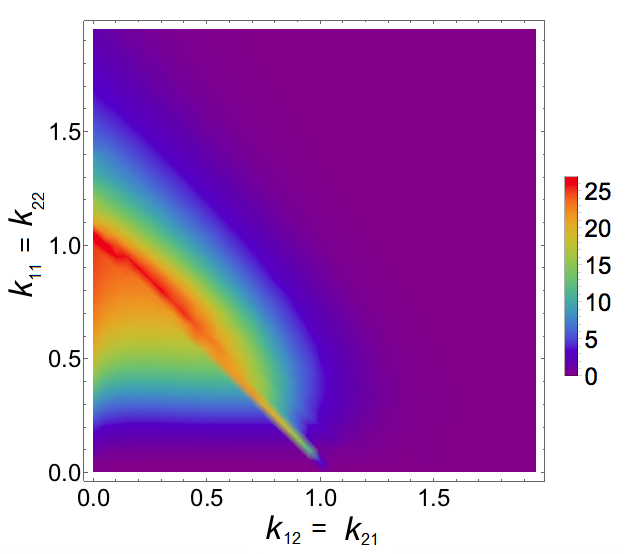}%
  \label{v10sum}
}

\subfloat[]{%
  \includegraphics[clip,width=0.3\columnwidth]{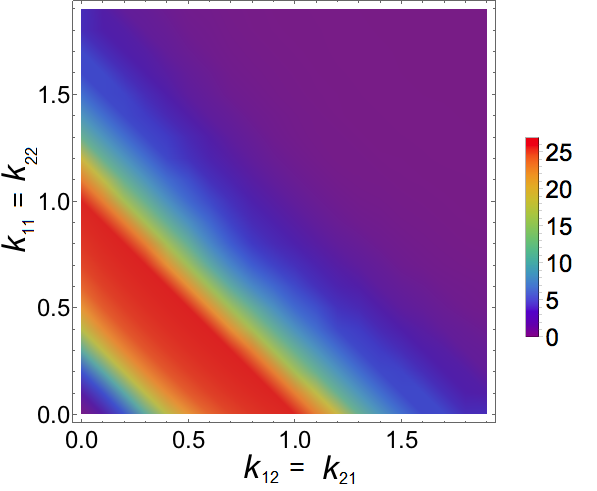}%
  \label{v10total}
}
\caption{The fitness landscapes plotted for rate constants $k_{21} = k_{12}$ and $k_{11} = k_{22}$ for (a) $\mathrm{MI_{sum}}$ and (b) $\mathrm{MI_{total}}$ and for system volume $V=10$. Phenomenologically, $V=10$ fitness landscapes are similar to $V=3$ (see Fig. 8 in main text).}
\label{landscape_large_V}
\end{center}
\end{figure}

\begin{figure}[H]
\begin{center}
\includegraphics[scale=0.55]{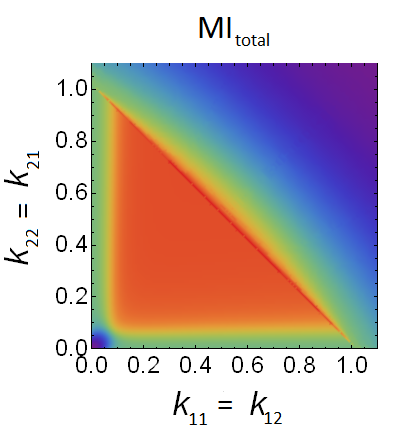}
\caption{The fitness landscape plotted for rate constants $k_{11} = k_{12}$ and $k_{22} = k_{21}$ for $\mathrm{MI_{total}}$ for $V=25$. The landscape shows similar qualitative behavior compared to low values of system volume $V$ (see Fig. 9 in main text).}
\label{v25Ring}
\end{center}
\end{figure}

In the main text, we used $V=3$ as the value for our system volume for the evolutionary simulations. Here we ensure that the qualitative behavior of the fitness landscapes remains similar for larger values of system volume $V$. Figs. \ref{v10sum} and \ref{v10total} show the fitness landscapes for  $\mathrm{MI_{total}}$ as a function of crosstalk versus direct rate constants; while the peaks in the landscapes broaden, the qualitative behavior does not change. Finally, Fig. \ref{v25Ring} shows the behavior of the landscape related to bifurcations for $V=25$ (see Fig. 9 in main text). 

\captionsetup[subfigure]{singlelinecheck=false,justification=raggedright,position=top}

%


\begin{thebibliography}{99}

\bibitem{Jordan}
J. D. Jordan, E. M. Landau, R. Iyengar,  \textit{Signaling networks: the origins of cellular multitasking}, Cell. \textbf{103}, 193 (2000).

\bibitem{crosstalk_experimental_exist_1}
S. M. Hill, \textit{Receptor crosstalk: Communication through cell signaling pathways}, Anat. Rec. \textbf{253}, 42 (1998).

\bibitem{crosstalk_experimental_exist_3}
J. Ptacek, \textit{et al}.,  \textit{Global analysis of protein phosphorylation in yeast}, Nature, \textbf{438}, 679 (2005).

\bibitem{crosstalk_experimental_exist_2}
J.A. Ubersax, \textit{et al}.,  \textit{Targets of the cyclin-dependent kinase Cdk1}, Nature, \textbf{425}, 859 (2003).

\bibitem{CT_experiment_4}
M. L. Schmitz, A. Weber, T. Roxlau, M. Gaestel, M. Kracht, \textit{Signal integration, crosstalk mechanisms and networks in the function of inflammatory cytokines}, Biochimica et Biophysica Acta (BBA) - Molecular Cell Research, \textbf{1813}, 2165 (2011).

\bibitem{Deeds-Eukaryote}
M. A Rowland, W. Fontana, E. J. Deeds, \textit{Crosstalk and Competition in Signaling Networks} Biophysical Journal, \textbf{103}, 2389 (2012).

\bibitem{ashok}
S. M. Lyons, A. Prasad \textit{Cross-talk and information transfer in mammalian and bacterial signaling}, PloS One \textbf{7}, e34488 (2012).

\bibitem{CT_example_1_IGFITGFBeta_Interaction}
R. H. Chen, Y. H. Su, R. L. Chuang, T.Y. Chang \textit{Suppression of transforming growth factor-beta-induced apoptosis through a phosphatidylinositol 3-kinase/Akt-dependent pathway}, Oncogene \textbf{17}, 1959 (1998).

\bibitem{CT_example_2_camp_mapk_interaction}
M. Saxena, S. Williams, K. Tasken, T. Mustelin, \textit{Crosstalk between cAMP-dependent kinase and MAP kinase through a protein tyrosine phosphatase} Nat. Cell Biol., \textbf{1}, 305 (1999).

\bibitem{CT_example_3_integrin_signaling}
M. A. Schwartz, M. H. Ginsberg, \textit{Networks and crosstalk: integrin signalling spreads} Nat. Cell Biol. \textbf{4}, E65 (2002).

\bibitem{CT_example_4_hunter}
T. Hunter,  \textit{The age of crosstalk: Phosphorylation, ubiquitination, and beyond}, Mol. Cell, \textbf{28}, 730 (2007).

\bibitem{CT_example_5_yan}
Y. Yan, C. L. Wei, W.R. Zhang, H. P. Cheng, J. Liu,  \textit{Cross-talk between calcium and reactive oxygen species signaling}, Acta. Pharmacol. Sin, \textbf{27}, 821 (2006).

\bibitem{CT_example_6_nishi}
H. Nishi, E. Demir, A.R. Panchenko, \textit{Crosstalk between signaling pathways provided by single and multiple protein phosphorylation sites}, J. Mol. Biol., \textbf{427}, 511 (2015).

\bibitem{Deeds-Prokaryote}
M. A Rowland, E. J. Deeds, \textit{Crosstalk and the evolution of specificity in two-component signaling}, PNAS, \textbf{111}, 5550 (2014).

\bibitem{specificity_example_1_Sharad}
M. N. McClean, A. Mody, J.R. Broach, S. Ramanathan, \textit{Cross-talk and decision making in MAP kinase pathways}, Nat. Gen., \textbf{39} 409 (2007).

\bibitem{specificity_example_2_Behar}
M. Behar, H. G. Dohlman, T. C. Elston, \textit{Kinetic insulation as an effective mechanism for achieving pathway specificity in intracellular signaling networks}, Proc. Natl. Acad. Sci. USA., \textbf{104}, 16146 (2007).

\bibitem{specificity_example_3_Bardwell}
L. Bardwell, \textit{Mechanisms of MAPK signalling specificity}, Biochem. Soc. Trans., \textbf{34}, 837 (2006).

\bibitem{specificity_example_4_Bardwell}
W. Kolch, \textit{Coordinating ERK/MAPK signalling through scaffolds and inhibitors}, Nat. Rev. Mol. Cell. Biol., \textbf{6}, 827 (2005).

 \bibitem{specificity_example_5_Flatauer}
L. J. Flatauer, S. F. Zadeh, L. Bardwell, \textit{Mitogen-Activated Protein Kinases with Distinct Requirements for Ste5 Scaffolding Influence Signaling Specificity in Saccharomyces cerevisiae}, Mol. Cell. Biol., \textbf{25} 1793 (2005).

\bibitem{specificity_example_6_Dard}
N. Dard, M. Peter, \textit{Scaffold proteins in MAP kinase signaling: more than simple passive activating platforms}, BioEssays, \textbf{28},146 (2006).

\bibitem{Muller}
R. Muller \textit{Crosstalk of Oncogenic and Prostanoid Signaling Pathways}, J. Cancer Res. Clin. Oncol., \textbf{130}, 429 (2004). 

\bibitem{Shi}
W. Shi, A. L. Harris, \textit{Notch signaling in breast cancer and tumor
angiogenesis: cross-talk and therapeutic potentials}, J Mammary Gland Biol. Neoplasia, \textbf{11}, 41 (2006).

\bibitem{Kalaitzidis}
D. Kalaitzidis, T.D. Gilmore, \textit{Transcription factor cross-talk: the estrogen receptor and NF-kappaB}, Trends Endocrinol. Metab., \textbf{16}, 46 (2005).

\bibitem{evolution}
S. J. Maynard, \textit{The Theory of Evolution}, (Cambridge University Press, Cambridge, England, 1993).

\bibitem{soyer}
 O. S. Soyer, S. Bonhoeffer, \textit{Evolution of complexity in signaling pathways}, Proc. Natl Acad. Sci. USA, \textbf{103}, 16337 (2006).

\bibitem{Mobashir}
 M. Mobashir, B. Schraven, T. Beyer, \textit{Simulated Evolution of Signal Transduction Networks}, PLoS ONE \textbf{7} e50905 (2012).

\bibitem{zulfikar}
M. Z. Ali, N. S. Wingreen, and R. Mukhopadhyay, \textit{Hidden long evolutionary memory in a model biochemical network},  arXiv:1706.08499 [q-bio.MN].

\bibitem{geneDup_Prince}
V. Prince, F. Pickett, \textit{Splitting pairs: the diverging fates of duplicated genes}, Nat. Rev. Genet. \textbf{3}, 827 (2002).

\bibitem{shannon}
C.E. Shannon, \textit{A Mathematical Theory of Communication}, The Bell System Technical Journal, \textbf{27}, 379 (1948).

\bibitem{Gillespie}
D. T. Gillespie, \textit{The chemical Langevin equation}, J. Chem. Phys., \textbf{113}, 297 (2000).

\bibitem{gonze} 
D. Gonze, A. Ouattara, \textit{Stochastic simulations Application to biomolecular networks},  (2014) 
\lq{homepages.ulb.ac.be/~dgonze/TEACHING/stochastic.pdf}\rq.

\bibitem{ladbury}
J.E. Ladbury, S.T. Arold, \textit{Noise in cellular signaling pathways: causes and effects}, Trends Biochem. Sci. \textbf{37}, 173 (2012).

\bibitem{moran} 
P. A. P. Moran, \textit{Random processes in genetics}, Proc. of the Cambridge Philosophical Society, \textbf{54}, 60 (1958).

\bibitem{nowak}
M. A. Nowak, \textit{Evolutionary Dynamics: Exploring the Equations of Life}, Belknap Press, (2006).

\bibitem{Risken}
H. Risken, \textit{The Fokker-Planck Equation}, Springer-Verlag Berlin Heidelberg (1984).

\bibitem{vanKampen} 
N.G. Van Kampen, \textit{Stochastic Processes in Physics and Chemistry}, 3rd Edition, North Holland (2007). 

\bibitem{supp}
See Supplemental Material at [] for calculation details and additional information.

\bibitem{Garcia-Palacios}
 J.L. Garcia-Palacios, \textit{Introduction to the theory of stochastic processes and Brownian motion problems}, arXiv:cond-mat/0701242.

\bibitem{multivariateFP}
Gillespie, D. T. \textit{The multivariate Langevin and Fokker-Planck equations}, Am. J. Phys., \textbf{64}, 1246 (1996).

\bibitem{papin}
J. A. Papin, B. O. Palsson, \textit{Topological analysis of mass-balanced signaling networks: a framework to obtain network properties including crosstalk}, J. Theor. Biol., \textbf{227}, 283 (2004).

\bibitem{chen_pleiotropic_1}
Y. Z. Chen, J. Qiu, \textit{Pleiotropic Signaling Pathways in Rapid, Nongenomic Action of Glucocorticoid}, Mol. Cell. Biol. Res. Commun., \textbf{2}, 145 (1999).

\bibitem{Pagliari_pleiotropic_2}
S. Pagliari, J. Jelinek, G. Grassi, G. Forte, \textit{Targeting pleiotropic signaling pathways to control adult cardiac stem cell fate and function}, Frontiers in physiology, \textbf{5}, 219 (2014).

\bibitem{Granek_pleiotropic_3}
J. A. Granek, O. Kayikci, P. M. Magwene, \textit{Pleiotropic signaling pathways orchestrate yeast development}, Current opinion in microbiology, \textbf{14}, 676 (2011).

\bibitem{gustin}
M. C. Gustin, J. Albertyn, M. Alexander, K. Davenport, \textit{MAP kinase pathways in the yeast Saccharomyces cerevisiae}, Microbiol. Mol. Biol. Rev., \textbf{62}, 1264 (1998).









\bibitem{telatar}
I. E. Telatar,  \textit{Capacity of Multi-Antenna Gaussian Channels}, Tech. Rep. Bell Labs, Lucent Technologies, (1995).

\bibitem{Raser}
J.M. Raser , E.K. O'Shea, \textit{Noise in Gene Expression: Origins, Consequences, and Control}, Science, \textbf{309}, 2010 (2005).

\bibitem{jangir}
J. Selimkhanov, \textit{et al}. \textit{Accurate information transmission through dynamic biochemical signaling networks}, Science \textbf{46}, 1370 (2014).

\bibitem{luca}
L. Cardelli, A. Csikasz-Nagy, N. Dalchau, M. Tribastone, M. Tschaikowski, \textit{Noise reduction in complex biological switches. Scientific Reports}, \textbf{6}, 20214 (2016).

\bibitem{swain}
P. S. Swain, M.B. Elowitz,  E. D. Siggia, \textit{Intrinsic and extrinsic contributions to stochasticity in gene expression}, Proc. natl. Acad. Sci. USA, \textbf{99}, 12795 (2002).

\bibitem{attisano}
L. Attisano , E. Labbe,  \textit{TGFbeta and Wnt pathway cross-talk}, Cancer Metastasis Rev.,  \textbf{23}, 53 (2004).

\bibitem{sumi}
T. Sumi, N. Tsuneyoshi, N. Nakatsuji, H. Suemori, \textit{Defining early lineage specification of human embryonic stem cells by the orchestrated balance of canonical Wnt/{beta}-catenin, Ac-tivin/Nodal and BMP signaling}, Development,  \textbf{135}, 2969 (2008).

\bibitem{ned_quorum}
T. Long , K.C. Tu , Y. Wang, P. Mehta, N. P. Ong, \textit{et al}., \textit{Quantifying the Integration of Quorum-Sensing Signals with Single-Cell Resolution}, PLoS Biology, \textbf{7}, e1000068 (2009).

\bibitem{madhani}
M. A. Schwartz, H.D. Madhani, \textit{Principles of MAP kinase signaling specificity in Saccharomyces cerevisiae}, Annu. Rev. Genet., \textbf{38}, 725 (2004).

\bibitem{dohlman}
Dohlman HG (2002) Annu Rev Physiol 64:129-152
H. G. Dohlman, \textit{G proteins and pheromone signaling}, Annual Review of Physiology, \textbf{64},129 (2002).

\bibitem{kramer}
B. D. Gomperts, P. E. R. Tatham, I. M. Kramer, \textit{Signal transduction} 1st Edition Amsterdam Elsevier Academic Press, (2004).

\bibitem{wolanin}
 P. M. Wolanin, P.A. Thomason, J. B. Stock, \textit{Histidine protein kinases: key signal transducers outside the animal kingdom}, Genome Biology \textbf{3}, 3013.1 (2002).

\end{thebibliography}

\begin{thebibliography}{9}

\bibitem{risken}
H. Risken, \textit{The Fokker-Planck Equation}, Springer-Verlag Berlin Heidelberg (1984).

\bibitem{vanKampen} 
N.G. Van Kampen, \textit{Stochastic Processes in Physics and Chemistry}, 3rd Edition, North Holland (2007).  
 
\bibitem{Garcia-Palacios}
 J.L. Garcia-Palacios, \textit{Introduction to the theory of stochastic processes and Brownian motion problems}, arXiv:cond-mat/0701242.
 
\bibitem{zulfikar1}
M. Z. Ali, N. S. Wingreen, and R. Mukhopadhyay, \textit{Hidden long evolutionary memory in a model biochemical network},  arXiv:1706.08499 [q-bio.MN].

\end{thebibliography}
\end{document}